# Collapse of Probability Distributions in Relativistic Spacetime

**Hans C. Ohanian**
(hohanian@uvm.edu)

*Department of Physics*
*University of Vermont, Burlington, VT 05405-0125*

*February 28, 2017*

**Abstract.** The collapse of a spatial probability distribution—such as the probability distribution associated with a quantum-mechanical wavefunction—is triggered by a measurement at a given spacetime point. It is customarily assumed that this collapse occurs along an equal-time hypersurface, say, $t = 0$, in whatever reference frame is being used. However, such a naïve instantaneous collapse process is inconsistent with relativity, because the equal-time hypersurfaces of different inertial reference frames are different. The attempts at implementation of instantaneous collapse in several different reference frames then lead to violations of probability conservation and violations of the scalar character of the probability contained in given volume elements. This problem affects not only the Copenhagen interpretation of quantum mechanics, but also other interpretations—such as the many-worlds interpretation and the consistent-histories interpretation—in which it is still necessary to specify what changes in probabilities occur when and where in a manner consistent with relativistic spacetime geometry. In the 1980s Schlieder and Hellwig and Kraus proposed that collapse of the probability distribution along the past light cone of the measurement point avoids these difficulties and leads to a Lorentz-invariant collapse scenario. Their proposal received little attention and some negative criticisms. In this paper I argue that the proposed past-light cone collapse is not only reasonable, but is compelled by Lorentz invariance of probability conservation, and is equally valid for the spatial probability distributions in quantum mechanics and for those in a game of chance, for instance, the probability distribution for a game with playing cards scattered over some spatial region. I examine the objections that have been made to the past-light-cone collapse scenario and show that these objections are not valid. Finally, I propose two possible interferometer experiments that can serve as direct tests of past-light-cone collapse, one with an atom interferometer, and the other with a light interferometer.

## I. Introduction

In contrast to the unquestionable success of quantum mechanics as a theory of the dynamics of atomic and subatomic particles, the fundamental meaning of the quantum-mechanical state vectors and wavefunctions and the interpretation of the effects of measurements remain controversial and murky. After almost a hundred years of investigations by physicists and philosophers, the interpretation problem is still mired in bizarre puzzles and paradoxes, with the orthodox Copenhagen interpretation under siege by dozens of new interpretation schemes, each staunchly defended by its partisans and vigorously attacked by the partisans of other schemes. These catfights have often produced more heat than light, but they have given us a better appreciation and respect for the unexpected difficulties that still lie hidden in the shadowy depths of quantum mechanics.[1]

Quantum mechanics suffers from what Wigner, in his judicious 1976 lectures on the interpretation problem, called the quantum-measurement paradox, that is, “the contradiction between the deterministic nature of the quantum-mechanical equations of motion and the probabilistic outcome of the measurements.” Wigner proposed to resolve this paradox by assigning to the state vector the role of a mere calculational tool instead of a representation of physical reality: "the quantum-mechanical equations of motion do not describe the measurement process; they only help in the calculation of the probabilities for different outcomes...The formalism of state vectors, equations of motion, etc., are only means to calculate these probabilities…The state vector does not represent 'reality'. It is a calculational tool." [2]

Wigner thought that his view of the role of the state vector was in accordance with the view that von Neumann expressed in his book on the *Mathematical Foundations of Quantum Mechanics,* where he used the state vector as a mathematical tool to formulate the two processes for changes of probabilities in ensembles of quantum-mechanical systems: Process 1, consisting of "arbitrary changes," or collapses, of the state vector, triggered by measurements with probabilistic outcomes according to the Born rule; and Process 2, consisting of the "automatic changes" that occur with the passage of time according to the dynamical equations of quantum mechanics.[3]

Expressed mathematically, Process 1 results in a sudden change $|\psi\rangle \rightarrow |\phi_n\rangle$ with probability $|\langle\phi_n|\psi\rangle|^2$, where the pre-collapse state vector is denoted by $|\psi\rangle$ and the complete set of eigenvectors for the possible outcomes of the measurement is denoted by $|\phi_n\rangle$, $n=1,2,3,...$ Process 2 results in a smooth gradual evolution according to the Schrödinger equation, $d|\psi\rangle/dt=(1/i\hbar)H|\psi\rangle$. Because von Neumann wanted to emphasize that these changes apply to statistical ensembles, he actually expressed these two processes as changes in the density matrix *U* (which might be that of a pure state or that of a mixture, $U=|\psi\rangle\langle\psi|$ or $U=\sum w_n|\chi_n\rangle\langle\chi_n|$ with probability weights $w_n$ for a set of states $|\chi_n\rangle$ ).Wigner felt that his own proposed demotion of the role of the state vector was in accord with von Neumann's focus on such changes in statistical quantities.

But Wigner's (or von Neumann's) proposed resolution of the measurement paradox amounts to no more than a consecration of the dichotomy between the two processes for changes of probabilities, and Wigner's view of the demoted role of the state vector raises awkward metaphysical questions of what is "reality" and what is a faithful representation of reality in contrast to a mere "calculational tool." Aren't the generalized canonical coordinates of classical Hamiltonian mechanics also calculational tools, and therefore without any direct physical significance?

The troublesome Process 1 would become superfluous if we were to adopt the Everett many-worlds interpretation, in which there is no collapse of the state vector.[4] Instead, the interactions occurring during the measurement cause an entanglement of the state vectors of the quantum system with those of the observer, and the net result is that the system-observer state becomes a superposition of all possible joint system-observer state vectors of the form $|\phi_n\rangle|\textbf{observer perceives } \phi_n\rangle$, with concordance between system state and observer state. In this context, the "observer" need not be human; instead it could be an automaton equipped with processing devices for the raw data arriving from external sensors and with macroscopic memory devices for permanent archival storage of the processed data; "perception" occurs when the data has been processed and is ready for insertion into the memory.[5] Thus the measurement causes the world to split into many worlds, or, more precisely, into many branches of the world, and in subsequent measurements each branch will branch out again and again. Everett assumed that each branch forever remains orthogonal to all the others and does not interact with any of them, so all branches evolve independently. Wigner dismissed this interpretation peremptorily, saying "From a positivistic point of view, the statement that there are such worlds, and that they are constantly created in large numbers, is entirely meaningless." [6]

An obvious deficiency of von Neumann's two-process prescription is the lack of any precise instructions of when and where Process 2 is to be halted and Process 1 is to be implemented. Does the evolution of the cat in Schrödinger's cat-in-a-box Gedankenexperiment proceed according to Process 1 or 2? The decoherence interpretation of quantum mechanics developed since the 1980s sheds some light on this question by showing how external perturbations from the environment can contribute random deviations to the phases of the quantum-mechanical state vectors, so interference terms, or cross terms, are effectively eliminated in the calculation of expectation values (stated otherwise, the off-diagonal terms in the density matrix are eliminated).[7, 8, 9, 10, 11] The probability distribution then mimics a non-quantum probability distribution, as in a game of chance. And, as in a game of chance, an experimenter who analyzes statistical data from a series of repetitions of the experiment will conclude that the outcome of the experiment is indeed probabilistic.

The further development of this interpretation led to the recognition that scattering of the pervasive thermal radiation found in all laboratory environments disturbs the quantum system being measured and quickly leads to decoherence for systems of macroscopic size, which explains why such systems never exhibit interference effects. Supporters of decoherence have enthusiastically declared that such an eradication of interference solves the measurement paradox, and it is indeed true that decoherence is crucial for guaranteeing an effective orthogonality and an absence of interference between different macroscopic states, so the superpositions of apparatus states generated by the system-apparatus interactions are saved from the disastrous syndrome of the Schrödinger cat paradox.[12, 13]

However, whether the decoherence interpretation actually saves us from the measurement paradox still remains controversial. Thus, Adler argues that regardless of how good decoherence is at eliminating interference among macroscopic states, it still leaves us with a probabilistic combination of several states instead of the single state an experimenter actually perceives.[14]

Another interpretation that has gained considerable support is the consistent histories (also called decoherent histories) formulation of Griffiths.[15] This is a prescription that tells us to start with a given initial state and then calculate the probabilities for a specified historical time sequence of possible outcomes of measurements. In this calculation the state vector is treated as a tool for constructing projection operators $|\varphi_n\rangle\langle\varphi_n|$ at the specified times that characterize the individual epochs in the history, from one measurement to the next. With these projections operators we can calculate the statistical correlations between the outcomes of the measurements at the ends of the epochs in the history. To some extent, this prescription and its reliance on projection operators was already anticipated in Wigner's lectures.[16]

Others schemes—such as Bohm's hidden-variables theory, also called de Broglie-Bohm mechanics or Bohmian mechanics,[17] and Ghirardi-Rimini-Weber's unified dynamics,[18] which is based on spontaneous stochastic collapses—are not actually reinterpretations of the standard formalism of quantum mechanics, but radical modifications of the equations of motion for quantum particles with fundamentally different dynamics. These schemes lack acceptable relativistic generalizations, and on those grounds I here reject them as hors de concours.[19]

Weinberg, in his 2013 textbook *Lectures on Quantum Mechanics*, gives a brief but incisive critique of the present status of the interpretation problem, including the post-Wigner developments, such as decoherence and Griffiths consistent (and decoherent) histories. Weinberg emphasizes that in the Griffiths interpretation decoherence is needed to define the epochs in the histories, that is, these epochs are not set by arbitrary deadlines chosen at whim, but by when and where an experimenter intercepts the quantum system by an apparatus or some other macroscopic system that triggers decoherence; and that in the Everett interpretation decoherence is needed to establish an effective orthogonality between the different macroscopic states, or many-worlds. But beyond emphasizing decoherence, Weinberg does not commit himself to any particular interpretation, and he faults all of them for failing to produce a satisfactory proof of the Born rule. Weinberg's view of the measurement problem echoes that of Wigner: "…the state vector serves only as a predictor of probabilities, not as a complete description of a physical system," and he ends on a gloomy note, saying "My own conclusion (not universally shared) is that today there is no interpretation of quantum mechanics that does not have serious flaws, and we ought to take seriously the possibility of some more satisfactory other theory, to which quantum mechanics is merely a good approximation." [20]

In view of the detailed descriptions and solid critiques that Wigner, Weinberg and others have offered on the various interpretations of quantum mechanics, I will refrain from any further comments on the interpretation controversy. Instead, in this paper I will address a question that all the interpretations have in common, but that has almost universally been neglected: How must the collapse of a probability distribution that extends over a region of spacetime be modified for consistency with the relativistic spacetime geometry? [21]

For quantum systems that extend over a spacetime region and, accordingly, involve wavefunctions and probability distributions that extend over this region, the difference between the Newtonian and the relativistic spacetime geometries has drastic implications for the collapse. The required modifications will become significant whenever the spatial dimension of the region divided by the speed of light is of the same order of magnitude or larger than the characteristic time interval for completion of the measurement process. Such modifications need to be considered even in interpretations—for instance, the Everett many-worlds interpretation or the Griffiths consistent-histories interpretation—that avoid explicit mention of a collapse. In these interpretations we still have discrete instants of discontinuity—that is, discrete spacelike hypersurfaces of discontinuity—where a history suddenly changes or the world suddenly branches out into several alternative worlds, and such instantaneous changes require different treatment in Newtonian and in relativistic spacetime.

Possible "oddities" in the relativistic collapse of quantum-mechanical wavefunctions were first discussed by Bloch[22] who pointed out that instantaneous collapse along a $t$ = constant hypersurface would lead to violations of probability conservation. He contemplated, and then rejected, the possibility that such violations could be avoided by adopting a collapse hypersurface that coincides with the light cone of the detector point.

Schlieder[23] re-examined the relativistic collapse problem and established that the appropriate collapse hypersurface is the past light cone of the spacetime point at which the measurement is performed. Schlieder's proof relies on conservation of probability and on Lorentz invariance of the probability weights. His proof was intended for quantum-mechanical probability distributions, such as those involved in the famous Einstein-Rosen-Podolsky paradox. He failed to see that quantum mechanics is not required for his proof—Lorentz invariance or, alternatively, general coordinate invariance is sufficient. As we will see, collapse along the past light cone is equally valid for non-quantum probability distributions, such as the probability distributions that might arise from a game of chance. Furthermore, the high degree of symmetry of the light cone associated with Lorentz invariance is irrelevant—the deformed light cones we encounter in general relativity will serve equally well. Collapse of probability distributions along the past light cone is ultimately rooted in the topology of spacetime, and it would be valid even in a conformal Weyl spacetime, which has light cones but no metric geometry at all.

Schlieder's proposal for collapse along the past light cone was adopted by Hellwig and Kraus[24] for the analysis of measurements in local quantum field theories. Unfortunately, Hellwig and Kraus expressed their analysis in terms of field operators in the Heisenberg representation, which makes it difficult to grasp the intuitive content of their collapse formalism. Also, they did not fully explore the consequences of multiple sequential collapses, which can lead to rather surprising results.

The work of Schlieder and of Hellwig-Kraus has only rarely been mentioned in the voluminous literature on the interpretation of measurements in quantum mechanics (despite all their profound expertise in relativistic quantum mechanics, neither Wigner nor Weinberg discuss this or any other aspect of relativistic collapse). Aharonov and Albert[25] briefly mentioned the Hellwig-Kraus treatment of collapse and quickly rejected it. They suggested that it would be better to insist on instantaneous collapse in some preferential reference frame and accept the ensuing violation of probability conservation and violation of Lorentz invariance of the collapse. They even suggested that collapse of a quantum-mechanical probability distribution might lead to different numbers of particles observed in different reference frames (!).[26]

The antirelativistic collapse schemes proposed by Bloch and by Aharonov and Albert—which rely on the adoption of a preferential reference frame—imply violations of fundamental tenets of relativity and quantum mechanics, such as conservation and Lorentz invariance of probabilities, conservation and Lorentz invariance of various quantum numbers that are directly proportional to particle numbers (baryon number, lepton numbers, strangeness, etc.), unitarity and Lorentz invariance of the S-matrix elements (which rest on the transformation laws of probabilities), simple Lorentz transformation laws for wavefunctions in accordance with Wigner's classification of irreducible representations of the Lorentz group (scalars, vectors, spinors, etc.), and the customary reliance on Lorentz invariance in the construction of Lagrangians and conservation laws for the fundamental interactions of high-energy physics. Given the absence of any empirical evidence for such draconian modifications of relativistic quantum physics, none of this seems plausible.

As a prelude to the formulation of a simple and consistent scheme for relativistic collapse of quantum-mechanical wavefunctions and probability distributions, Sections II and III of this paper discuss a simple example of relativistic collapse of the probability distribution in a game of chance involving a random spatial distribution of playing cards. This example directly leads to the conclusion that the collapse must proceed along the past light cone of the detection point of a card. Furthermore, if two or more of the light cones belonging to different detection points intersect (that is, are not nested one within the other), the collapse proceeds along the boundary of the topological union of these light cones, and collapse along such a hypersurface, with peaks and valleys, can lead to a curious “pre-collapse” at a time earlier than the detections. Section IV then applies the lessons learned from the game of chance to the quantum-mechanical case, and Section V applies them to the EPR experiment and “delayed-choice” experiments with photons, such as Aspect’s experiments with coherent light, and explores the implications for the spooky action-at-a-distance (“spuckhafte Fernwirkung”) that Einstein regarded as an intolerable defect of quantum mechanics. Section VI outlines an experimental atom-interferometer test and a light-interferometer test that could verify the expected past-light-cone collapse. Finally, Section VII gives some general comments.

## II. Collapse of a probability distribution for a game of chance

To understand the collapse of a probability distribution in relativistic spacetime, it is helpful to begin with a simple example of a game of chance involving a spatial distribution of playing cards. Take a standard deck of 52 playing cards, seal the individual cards into 52 identical envelopes, and mail them at random to 52 recipients at locations all over the globe. The recipient who receives the Queen of Spades is the winner of this game. If we focus our attention on the Queen, the mailing generates a random probability distribution for the Queen extending over 52 spacetime locations, with a probability of 1/52 for each location (cards other than the Queen of Spades merely serve as “fillers” or “nulls” to give all sealed envelopes the same mechanical properties, so a perfectly random chaotic mixing of the envelopes can be achieved by, say, tumbling them in a rolling drum). Figure 1 shows a spacetime diagram for the resulting probability distribution, or “probability tree.”

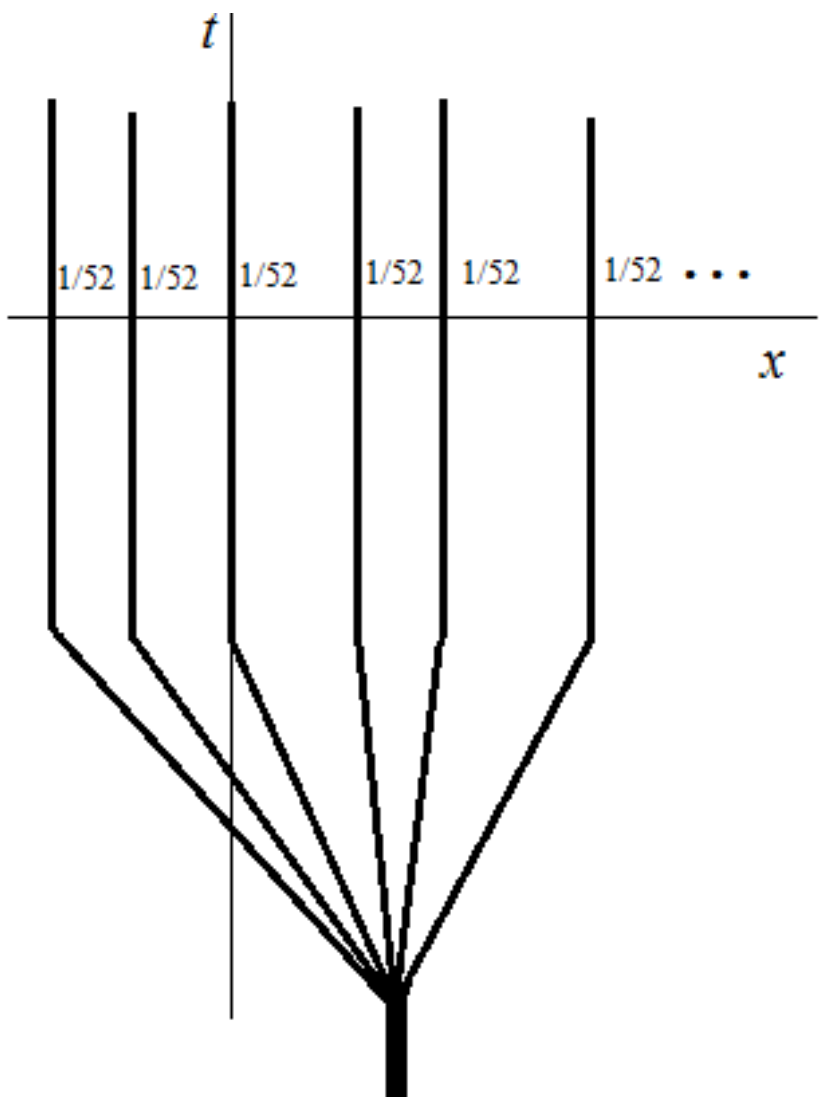

**Fig. 1** Spacetime diagram of the probability distribution for the Queen of Spades generated by mailing 52 playing cards at random to recipients at 52 locations. In this diagram, only a few of these locations are shown, all on the $x$ axis. The lines are the worldlines of the cards. After the cards reach their destinations, they are assumed to remain at rest (this assumption is merely made for the sake of simplicity; instead, we could assume that some or all of the cards, continue to move). Each of the worldlines has a probability weight of 1/52 for the Queen of Spades.

Whenever and wherever a recipient opens an envelope and inspects the card it contains, the probability distribution is altered. For instance, if the first opening at, say, $x = 0$, $t = 0$ does not reveal the Queen, the probability weight for that location collapses to 0, whereas it "instantly" increases from 1/52 to 1/51 at each of the other 51 locations. At some stage, the Queen will be found at some definite location, so the probability weight "instantly" increases to 1 for that location, and it collapses to 0 for all others—and the game ends. The total probability is conserved during every collapse, that is, the sum of probability weights for all the locations before the collapse is equal to 1, and the sum of probability weights after the collapse is also equal to 1.

In Newtonian spacetime, an instant of time is well defined, and it corresponds to a single, unique 3-D spatial hypersurface. There is then no objection to an instantaneous collapse of a probability distribution on this hypersurface. The collapse is invariant under the Galilean transformation between Newtonian inertial reference frames, with their absolute time. The invariance of the changes of the probability weights at different locations implies invariance of probability conservation, that is, the total probability is conserved in all Newtonian inertial reference frames .

However, in relativistic spacetime, time is not absolute, and an instant of time has no absolute meaning—two simultaneous events at different spatial locations in one inertial reference frame can occur at different times in another reference frame. Instantaneous collapses in two different relativistic inertial reference frames therefore lead to two different probability distributions and to violations of probability conservation, as illustrated in Fig. 2, which shows what happens to the probability distribution of Fig. 1 if it is subjected to instantaneous collapse in the $t, x$ reference frame and then examined in a different $t', x'$ reference frame. The discrepancy in the total sum of probability weights in the second reference implies that the naïve scheme of instantaneous collapse attempted in Fig. 2 is fatally flawed and must be rejected.

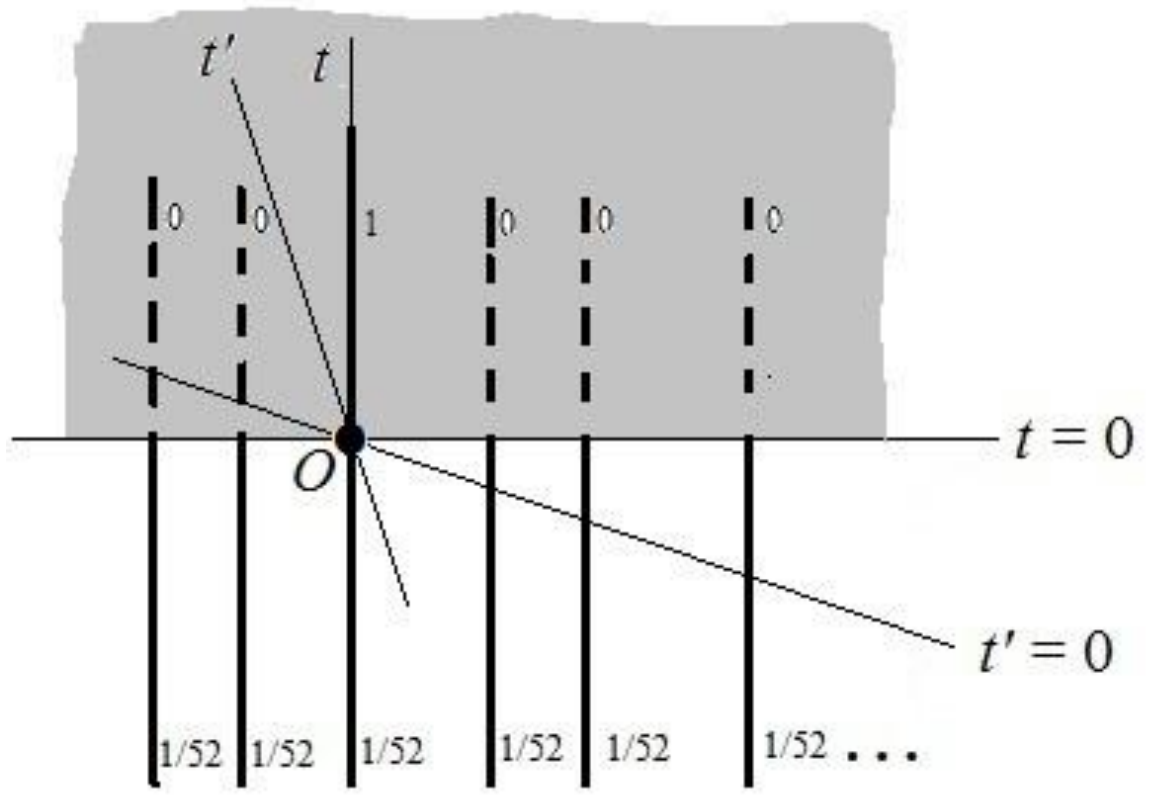

**Fig. 2** Parallel worldlines for the probability distribution of cards, and equal-time hypersurfaces $t = 0$ and $t' = 0$ in two inertial reference frames shown in a Minkowski diagram. If the collapse is instantaneous in the first reference frame, the inspection of the card at $t = 0,\ x = 0$ leads to a collapsed probability distribution for the gray spacetime region $t > 0$, whereas the white region $t < 0$ remains uncollapsed. If the Queen is found at the point $O$ (that is, at $t = 0,\ x = 0$), then in the second reference frame moving in the negative $x$ direction relative to the first, the net sum of probability weights evaluated along the hypersurface $t' = 0^+$ immediately after this instantaneous collapse is 1 + 1/52 + 1/52 + 1/52 +..., where all the worldlines with positive $x$ coordinates contribute a fraction 1/52 and those with negative $x$ coordinates contribute zero.

Another way to recognize the inconsistency between instantaneous collapse and relativity is by checking the relativistic transformation law for the speed of collapse. Instantaneous collapse in a given reference frame—say, the $t, x$ frame—implies that the collapse proceeds at an infinite speed $dx / dt = \infty$ along the $t =$ constant hypersurface. But in the $t', x'$ frame moving at speed $V$ in the negative $x$ direction (as in Fig. 2), the speed of collapse would then be finite, $dx'/dt' = (\infty + V)/(1+\infty V/c^2) = c^2/V$ . This obvious failure of the Lorentz invariance of infinite speed immediately suggests that we should adopt a speed of collapse equal to the speed of light, which we know to be invariant. And, indeed, we will see from further exploration of the collapse problem that the speed of light is the correct choice.

To formulate a new relativistically acceptable collapse scenario we first need a clearer understanding of the Lorentz-transformation properties of probabilities and their conservation law. For that purpose it is helpful to rely on an analogy with the well-known Lorentz-transformation properties of electric charge and its conservation, which can be derived from the differential conservation law $\partial\rho/\partial t + \nabla\bullet\mathbf{j} = 0$ , or its covariant form $\partial j^\mu/\partial x^\mu = 0$ .[27] To exploit the analogy between an electric charge distribution and the probability distribution, imagine that instead of just 52 playing cards we have a distribution of very many cards, so we can speak of a probability density for the cards and, if the cards are in motion, a probability current density.[28]

Integrating the differential conservation law $\partial j^\mu/\partial x^\mu = 0$ over a 4-D volume bounded by two parallel flat constant-time hypersurfaces $t = t_1$ and $t = t_2$ , and applying Gauss's theorem, we immediately find that $\int \rho d^3x$ is the same on both hypersurfaces, which expresses the conservation of the total probability calculated for successive instants of time. And integrating the differential conservation law over a 4-D volume bounded by two intersecting flat constant-time hypersurfaces (say, $t = 0$ and $t' = 0$ ), in two different inertial reference frames, we find that $\int \rho d^3x = \int \rho' d^3x'$ , which establishes that the total probability is a scalar. If the probability density is concentrated on a discrete set of individual pointlike bodies then the integrated probability density for each body is a scalar, analogous to the scalar electric charge of individual particles. This refutes conjectures occasionally offered in the quantum-mechanical literature that probabilities in different inertial reference frames might differ.

The requirements of probability conservation and equality of the total probability in all inertial reference frames conflicts with the instantaneous collapse scenario tentatively proposed in Fig. 2. Because the trouble in Fig. 2 arises from penetration of the flat hypersurface $t' = 0$ into the white uncollapsed region in the lower right quadrant, we can repair this trouble by deforming the boundary of the gray collapse region downward, so it lies below each of the flat hypersurfaces $t = 0, t' = 0, t'' = 0, t''' = 0, \ldots$ for all conceivable inertial reference frames, for all conceivable magnitudes and directions of the velocities of the Lorentz transformations. Obviously, the boundary of the gray collapsed region then coincides with the past light cone of the collapse point *O*, as shown in Fig. 3, and the gray collapse region engulfs all of the absolute future and also all of the "ambiguous" region.[29] With this choice of collapse surface, the demand for probability conservation will be satisfied in all inertial reference frames. Note that the rule of thumb for the collapse is that all regions that can collapse will collapse. The only region that is saved from collapse is the absolute past, which is unalterable.

This is, in essence, a paraphrase of the argument given by Schlieder, with the important modification that we are here applying it to a non-quantum probability distribution, and we are thereby establishing that the argument is valid for any probability distribution in relativistic spacetime, quantum or not. Also note that instead of examining what happens in a multitude of inertial reference frames related by Lorentz transformations, we could more simply examine what happens in noninertial general reference frames, with curved spacelike equal-time hypersurfaces that pass through the point *O*. On and above every such curved hypersurface the probability must be conserved.[30] Because the lowest of all such curved hypersurfaces lies along the past light cone, conservation on all such hypersurfaces immediately implies that the gray collapsed region has the past light cone as its boundary.

The collapse of the probability distribution along the past light cone suggests a highly counterintuitive picture of a collapse backward in time at the speed of light from the point *O* toward the outermost edges of the probability distribution. But it is better to think of this collapse as forward in time from the outer edges of the probability distribution toward the point *O*.

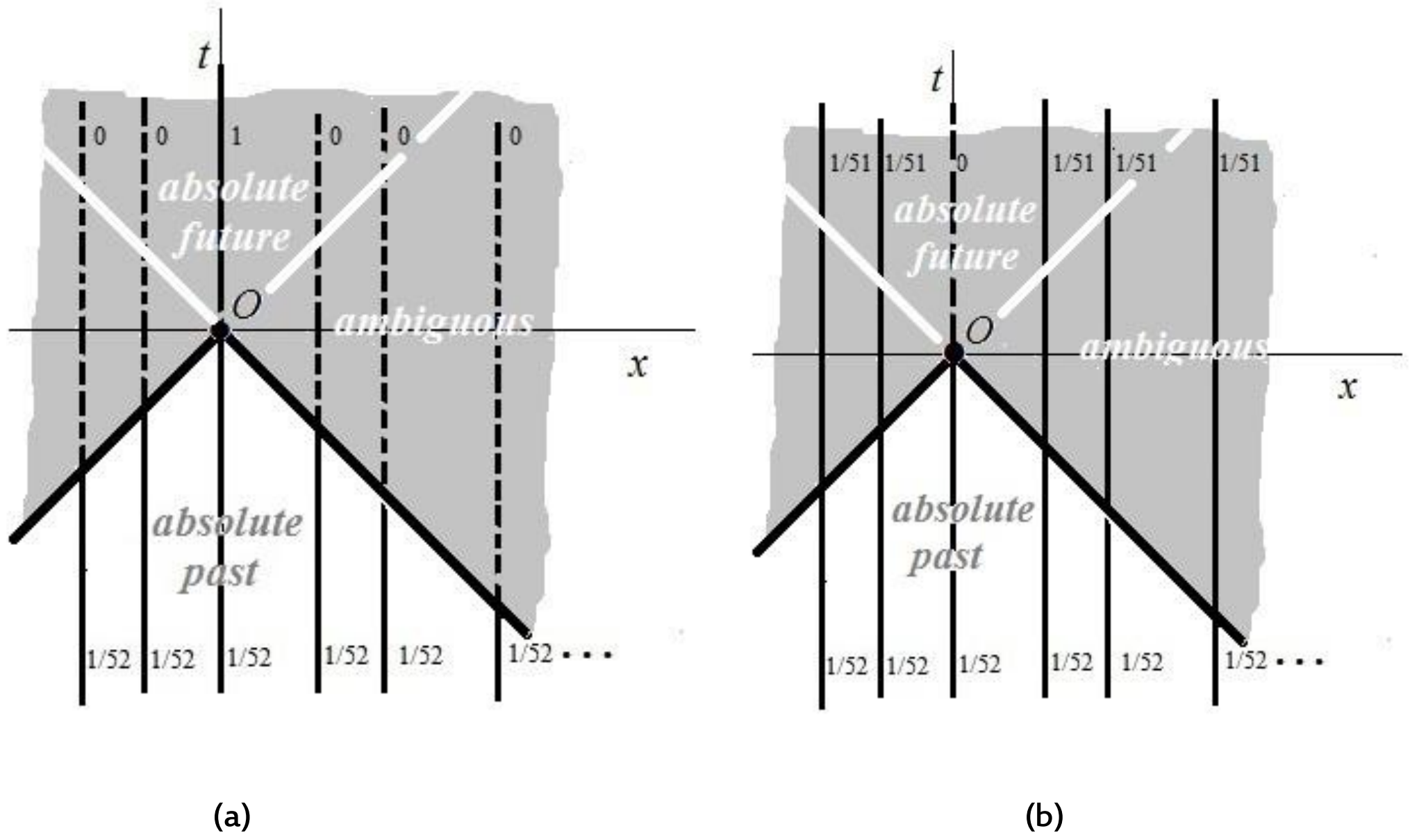

**Fig. 3** Relativistically, the boundary between the collapsed (gray) and uncollapsed (white) spacetime regions of the probability distribution coincides with the past light cone of *O*. The uncollapsed region is the absolute past of *O*, and the collapsed region is the topological union of the absolute future and the ambiguous region. (a) Uncollapsed and collapsed probability weights if the Queen is found at $x = 0$. (b) Uncollapsed and collapsed probability weights if the Queen is not found at $x = 0$.

The collapse along the past light cone is not produced by some spooky action-at-a-distance or some weird superluminal signals or disturbances. Collapse along the past light cone is merely a consistency requirement that we must impose on the collapse behavior of probability distributions for the sake of achieving invariance of probability conservation in all inertial reference frames. It should be viewed as analogous to Einstein's 1905 adoption of the equally weird and spooky postulate of a constant speed of light in all inertial reference frames, which he did for the sake of achieving invariance of clock synchronization procedures by light signals. Moreover, the relationship between Einstein's constant-speed-of-light postulate and past-light-cone collapse runs much deeper than an analogy: in the historical and logical development of physics, the past-light-cone collapse must be considered as a consequence of the constant speed of light incorporated in the construction of the theory of relativity and its Minkowski spacetime.

The relativistic collapse of a probability distribution along the past light cone is no more spooky than the instantaneous collapse in the Newtonian case, where the collapse at a remote location from the inspection point is also inferred from the conservation law. If you find the Queen of Spades at some location, you can infer it is absent from all other locations—no transmission of a physical signal of any kind is required, and nobody would think this in any way spooky or surprising. If the relativistic past-light-cone collapse seems somewhat strange at first sight, we have to learn to live with (as a last resort, we can console ourselves with von Neumann's dictum, "In mathematics we never understand anything, we just get used to it").

If we take the limit $c \to \infty$, the past light cone in Fig. 3 merges into the usual flat Newtonian $t = 0$ constant-time hypersurface, so the relativistic collapse of the probability distribution reduces to the expected Newtonian collapse at an instant of time. For a probability distribution of a size comparable with the size of the Earth, the time delay between the collapses at the most distant points of the probability distribution are of the order of a few hundredths of a second, so the relativistic collapse along the past light cone differs very little from the naïve instantaneous collapse expected in Newtonian spacetime.

Note that collapse *along* the past light cone does not mean collapse *on* the past light cone. The collapse cone lies in the ambiguous zone, infinitesimally close to the light cone, but not on light cone. The set of reference frames used in Schlieder's argument have speeds that asymptotically approach the speed of light, but never reach the speed of light (topologically, this set is an open set). And when we interpret the collapse diagram as indicating that the collapse happens earlier at large distance, this merely means "earlier" in the nominal sense, not in the absolute sense of a happening in our absolute past. The apparent pre-collapse at a remote location is an illusion resulting from the choice of an unsuitable time coordinate in the ambiguous region. We can always find a reference frame in which collapse at the spacetime point $O$ at the apex of the past light cone occurs before any selected spacetime point along (but just outside) the past light cone, in concordance with the causal chain that treats the inspection at $O$ as the cause of the consequent collapse of the probability distribution at the spacetime points along the light cone. For an observer evolving on her worldline at the fixed spatial point $x = 0$, the event $O$ enters her past light cone first (it becomes part of her observable world) and only then, an instant later, does the collapse enter her past light cone. The pre-collapse is an artifact of our choice of inertial reference frames with $t =$ constant hypersurfaces (a choice we make for pragmatic reasons, because it gives us a simple formulation for initial conditions and time evolution according to the hyperbolic differential equations that govern relativistic dynamics).

The time order of points along the past light cone is ambiguous and cannot be used for causal discrimination. This is quite analogous to what happens in the naïve Newtonian instantaneous collapse, where inspection at one point of the $t = 0$ hypersurface leads to collapse of the probability distribution at a remote point at the same instant. But if inspection and collapse occur at the same instant, then there is no clearly defined time order between these events, and we cannot assert that the time order determines what causes what. Causation must be determined by other means, such as the knowledge (or imagined knowledge) of the experimenter who made the decision to inspect at the selected time.

The past-light-cone collapse hypersurface is Lorentz invariant, and we could have tried to rely on this invariance requirement to discover the correct choice of hypersurface (in quantum mechanics, proposals for collapse on the past light cone have been based on this invariance requirement; see, e.g., Bloch[22]). However, reliance on nothing but Lorentz invariance is not quite sufficient to identify the collapse hypersurface, because it does not permit us to exclude collapse along an invariant hyperboloid [such as $t^2 - x^2 - y^2 - z^2 = m^2$, with some real number $m$] or collapse along the future light cone as an alternative to collapse along the past light cone.

A perceptive reader might complain that for the collapse along the past light cone shown in Fig. 3 probability conservation becomes problematic when the total sum of probability weights is reckoned along a $t =$ constant spacelike hypersurface that slices across the distribution somewhat below the apex of the past light cone, for which the sum of probability weights would be less than 1 (Fig. 3a) or more than 1 (Fig. 3b). However this is an erroneous way to reckon total probabilities. We will see later that the correct way is to use only those spacelike surfaces for summation or integration that avoid intersections with the collapse light cone, and these surfaces will in general have to be curved spacelike hypersurfaces, lying either entirely inside the collapse light cone or outside.

## III. Collapse topology

The uncollapsed spacetime region is the topological intersection $\bigcap$ of all the nominal past regions $t < 0, t' < 0, t'' < 0, \ldots$, and the collapsed region is the topological union $\bigcup$ of all the nominal future regions $t > 0, t' > 0, t'' > 0, \ldots$ These conclusions about the shape and location of the relativistic collapse surface are compelled by probability conservation and by the topology of the possible spacelike hypersurfaces and the topology of the light cone. Details of how the probabilities arise are irrelevant—the Schlieder argument applies equally well to probabilities in a game of chance and in quantum mechanics.

If in addition to the single inspection of a card at the spacetime point $O$ on one worldline we perform additional inspections of cards on other worldlines, we trigger more collapses of the probability distribution, until one of these inspections reveals the Queen, so the collapse process comes to its end. The simplest case of such a series of collapses involves spacetime points of inspection that have an absolute time order, with each spacetime point lying in the absolute past of the next spacetime point, and each past light cone lying inside the past light cone of the next spacetime point (see Fig. 4). The collapses along such "stacked" past light cones then occur in a clear and unambiguous time sequence, and each collapse merely modifies the results of the preceding collapse.

Inertial time coordinates and their flat equal-time hypersurfaces are not well adapted to collapses involving such successive collapses and their stacked past light cones. The successive light cones are the time boundaries of different worlds, each with a different probability distribution of cards. These different worlds (or, in the language of the Everett interpretation, different branches of our world) occupy different layers of spacetime, with a lower boundary consisting of one light cone and the upper boundary consisting of the next light cone (see Fig. 4). For systems of finite spatial size, the layers also have lateral boundaries, consisting of a cylindrical hypersurface surrounding the system. The equal-time hypersurfaces of an inertial reference frame cut across the boundaries between different worlds. This is harmless if the system's spatial boundaries lie within the boundaries of one world (see the hypersurface $t = t_1$ in Fig. 4), but it becomes problematic if the spatial boundaries lie beyond the boundaries of that world (see the hypersurface $t = t_2$) .

.

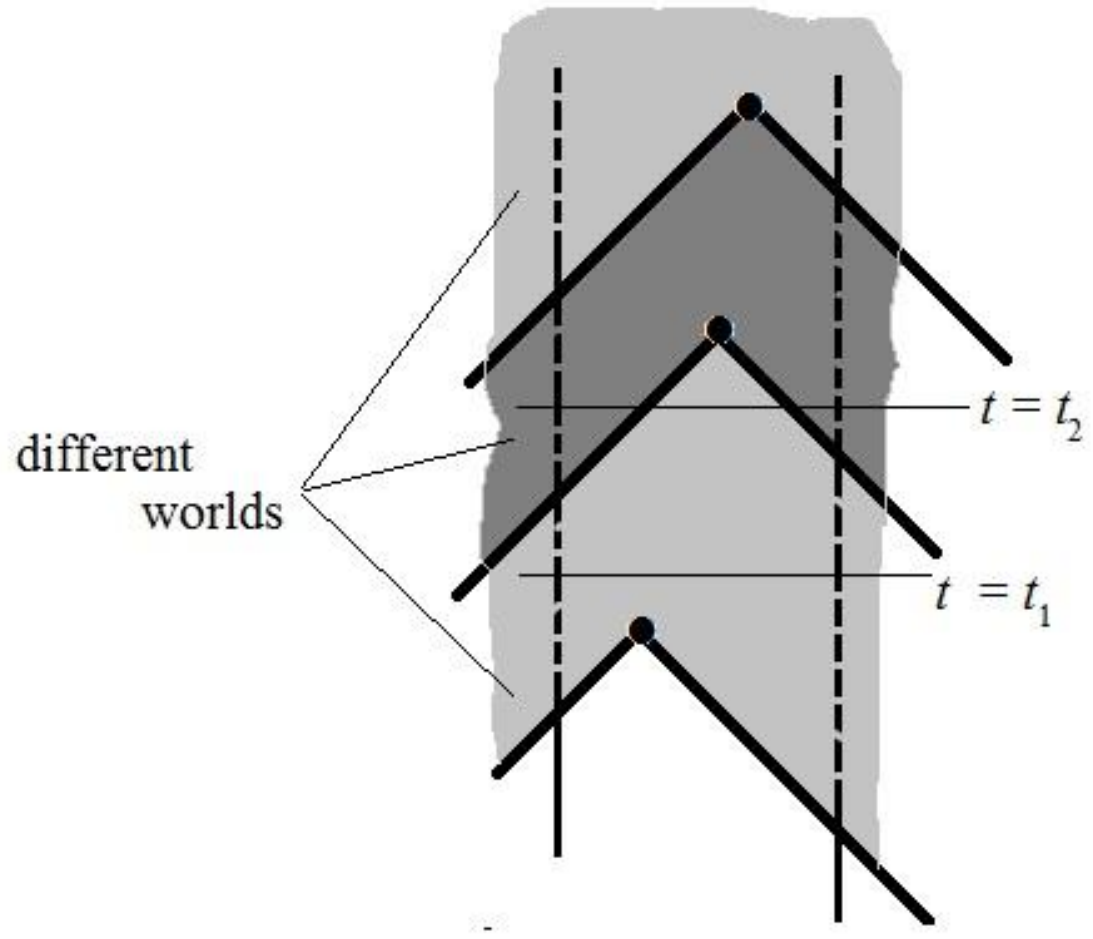

**Fig . 4** Successive collapses in a stack of light cones. Each of the layers of spacetime between two successive collapse light cones is occupied by a different collapsed probability distribution ("different worlds"). In the diagram, the different layers are shaded successively gray or dark gray. The distance between the vertical dashed lines indicates the maximum spatial extent of the system. The thin horizontal lines indicate two different hypersurfaces of constant time.

The layered structure of spacetime brought about by successive measurements is reminiscent of a layer cake with layers of different flavors. The discrete structure of such a stack of different worlds brought about by the succession of stochastic collapses of the probability distribution is in sharp contrast to the smooth structure of a world evolving according to deterministic classical mechanics, where all the changes arise from the dynamical equations of motion, and there is never any collapse process. In such a deterministic evolution, the flavors of the layers change gradually with the progress of time, and they meld one into another, without any sharp boundaries between the layers.

The example of successive collapse hypersurfaces shown in Fig. 4 is rather simple because each light cone lies within the absolute past or the absolute future of every other light cone. The slicing of spacetime into different worlds becomes far more complex if the successive collapse hypersurfaces are not stacked, with a well-defined time order, but intersect because each cone has a spacelike displacement relative to one or several of the other cones (see Fig. 5). We then cannot treat the collapses in succession, and instead must treat the collapses in conjunction.

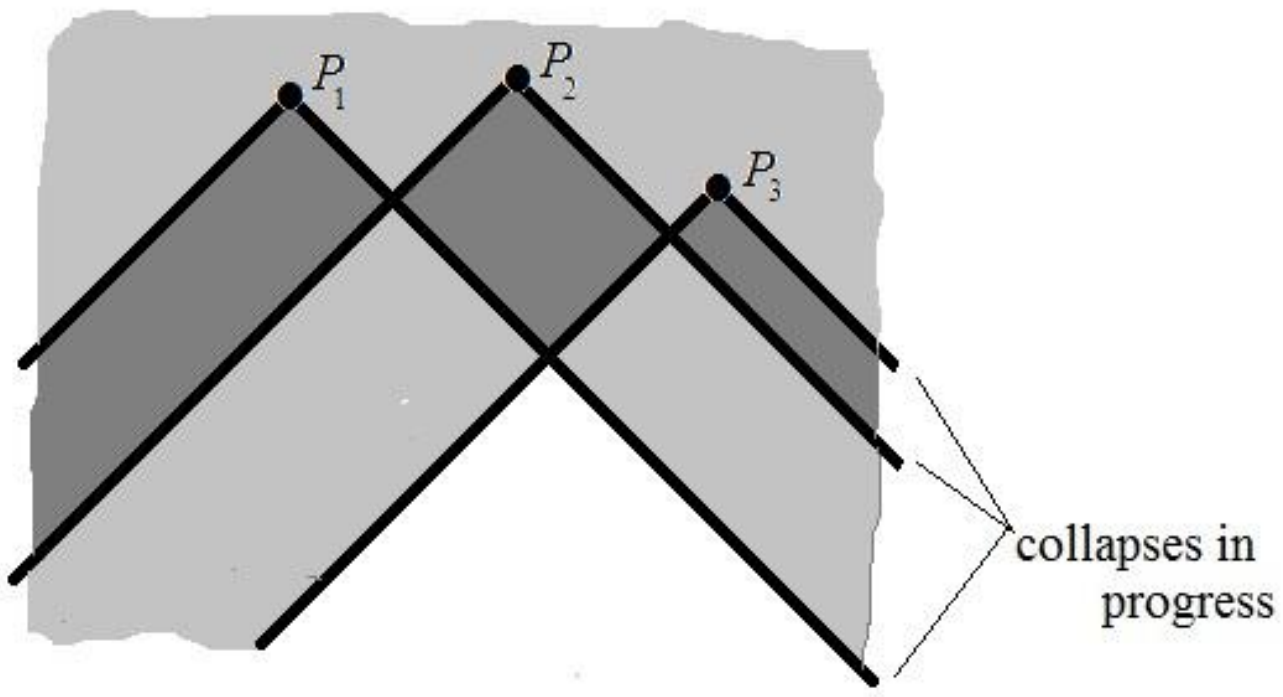

**Fig. 5** Collapses along several light cones with spacelike relative displacements of their apexes. The fully collapsed region (gray, at top) is now separated from the uncollapsed region (white, at bottom) by a complicated transitional region with intersecting past light cones and partial collapses (dark gray and gray, at center).

The uncollapsed region (white) is then the topological intersection of all the absolute past regions $I^-(P_i)$ contained in these light cones, $K^- = \bigcap_i I^-(P_i)$; and the fully collapsed region (gray) is the complement of the closure of the union of all the absolute past regions, $K^+ = M - \overline{\bigcup_i I^-(P_i)}$, where $M$ is the entire Minkowski spacetime manifold and the overbar indicates closure of an open set. The collapse begins on the surface of the region $K^-$, and it ends on the surface of $K^+$ (in topological notation, these two hypersurfaces are $\bar{K}^- \cap \left(\overline{M - K^-}\right)$ and $\bar{K}^+ \cap \left(\overline{M - K^+}\right)$). Topologically, these surfaces are cones, but not circular cones; their sides are segments of circular cones, with different vertices.

The region between these two surfaces (medium gray in Fig. 5) is a transitional region, in which the individual collapses are in progress, with a complicated fragmented cellular structure of different collapses in different cells of spacetime. Thus, we now have a layer cake in which the successive layers are separated by sheets of puff pastry (a veritable torte mille-feuilles).

To avoid excessive complications, it will be instructive to consider a game of chance with only two cards. This case is exceptional in that the complement rule for probabilities completely determines the outcome regardless of which card is inspected. If the two inspection points are separated by a spacelike displacement, then there exists an inertial reference frame in which they occur at the same instant of time, and, for convenience we adopt this reference time for our analysis. Accordingly, Fig. 6 shows the two worldlines and the two inspection points *A* and *B* at the same instant of time. Inspection at *A* triggers collapse along the past light cone of *A*. For instance, if the Queen is found at *A*, then the probability on the left worldline collapses to 1 above *A*, and the probability on the right worldline collapses to zero above *D* [see Fig. 6(a)]. If instead, we perform the inspection at *B*, then the probabilities above *C* and *B* collapse to 1 and zero, respectively [see Fig. 6(b)]. Furthermore, if we perform inspections at both *A* and at *B*, the probabilities on the segments *CA* and *DB* collapse to 1 and zero, respectively [Fig. 6(c)]. Thus, the values obtained for the probabilities on the segments *CA* and *DB* depend on where we make our inspection. Note that the probabilities of the uninspected worldlines in Figs. 6(a) and (b) are inferred from the complement rule, that is, by relying on probability conservation evaluated along spacelike hypersurfaces that pass just above the segments *DA* or *CB* of the light cone in Figs. 6(a) and (b), respectively.

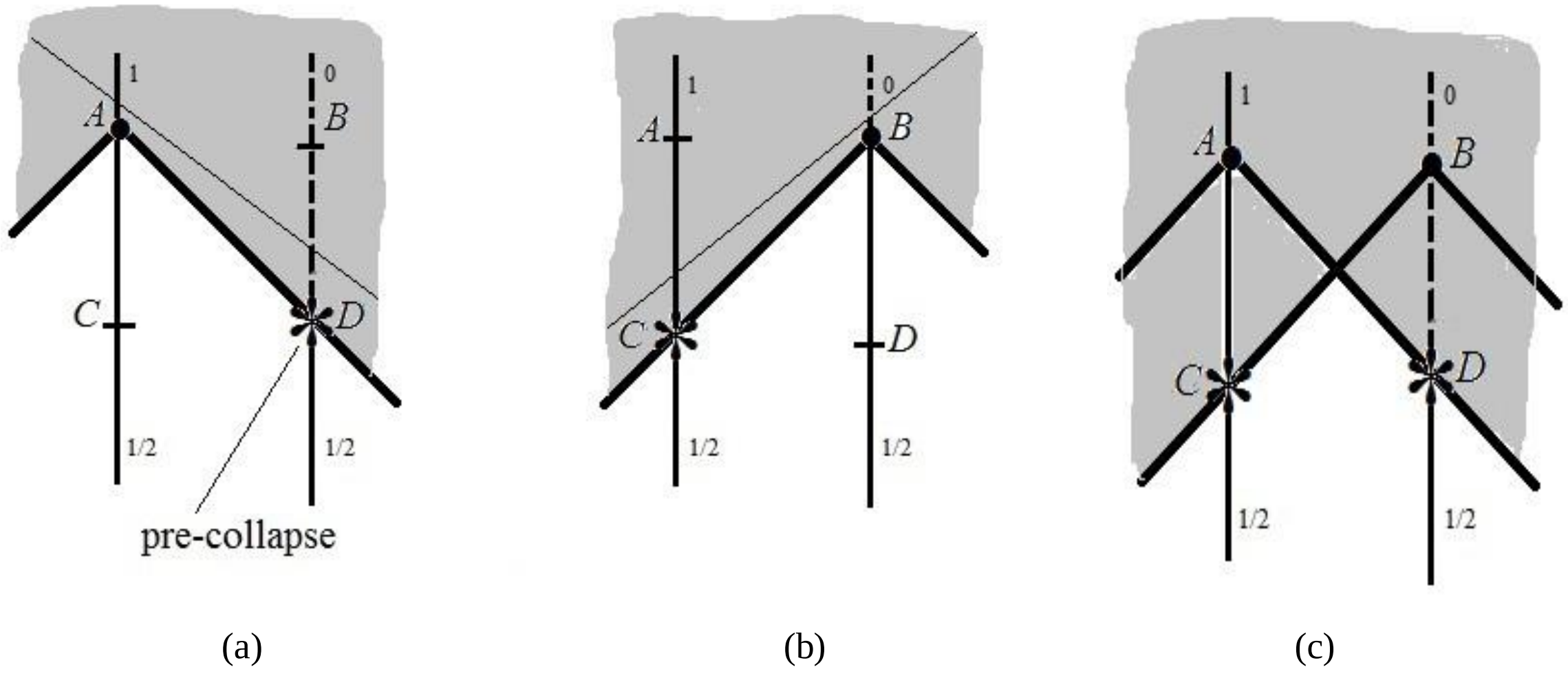

**Fig. 6** These diagrams show the two worldlines for two cards and the inspection points *A* and *B* in an inertial reference frame in which these two spacetime points occur at the same instant of time. The thin lines represent spacelike hypersurfaces used to evaluate, by the conservation law, the probabilities for the worldlines that have not been inspected. If there are the only two cards "in play," then the initial probabilities are ½ for each worldline. The numbers 1 and 0 indicate the final probability weights, after the collapse triggered by the inspections of cards. (a) Inspection at *A*. (b) Inspection at *B*. (c) Inspection at both *A* and *B*.

A puzzling aspect of such measurements at points separated by a spacelike displacement is the "pre-collapse" we see in Fig. 6, where the collapses of the probabilities at *C* and *D* on one or both of the two worldlines occurs at a time *earlier* than the inspections *A* and *B*. This pre-collapse is an instance of the spooky action-at-a-distance of the EPR experiment, now made even spookier by action into the past. The apparent reversal of the time order of cause and effect seems counterintuitive, but it leads to no detectable contradictions. As already mentioned in Section II, the collapse caused by the inspection at *A* (or at *B*) merely reaches into ambiguous region of *A* (or *B*, respectively), where the time order is not well defined, and we can construct inertial reference frames such that any selected spacetime point in the (open) set of points of the collapsed region near the past light cone is in the nominal past of the spacetime point *A* (or *B*) or in the nominal future, whatever we want. Therefore, we are not faced with an absolute reversal of time order. What is new in the double collapse arrangement shown in Fig. 6(c) is that the inspection at the spacetime point *A* reaches into the absolute past of the point *B* and viceversa. Each inspection therefore preempts the other, and each inspection can be regarded as a confirmation of the other (what in quantum mechanics is called a nondemolitional measurement).

If the transverse distance between *A* and *B* is about $10^7$ m (nearly antipodal points on the Earth), then the pre-collapse in the inertial $x$, $t$ reference frame of the Earth extends back in time by about 0.03 s. However, no message, in the form of a light pulse or particle pulse, can reach the points *A* or *B* from the collapsed region and thereby permit an experimenter at *A* or *B* to generate a contradiction by aborting the measurement operation that causes the collapse or to exploit the pre-collapse by laying bets on the outcome found at *A* and *B*.

It is also instructive to examine the conservation law for a time-dependent probability distribution. For this, we need to compare the sum of probability weights on two successive spacelike hypersurfaces cutting across the probability distribution. It is obvious that if both of these hypersurfaces are within one layer of the world (see the two surfaces $t = t_1$ and $t = t_3$ in Fig. 7), then the conservation law is valid. However, if one of these surfaces is partially in one world, and partially in another world with a different probability distribution (see $t = t_2$ in Fig. 7), then it seems that the conservation law will not be valid.

To ensure the validity of the conservation law for all spacelike hypersurfaces that slice across the spacetime diagram in Fig. 7, we must associate a probability weight not only with the particle worldlines in the system, but also with some or all of the light cones that intersect these worldlines. These light cones are surfaces of discontinuity for probability, and local validity of the conservation law requires a flow of probability along these surfaces from one worldline to neighboring worldlines, so as to preserve the total probability.[31] Figure 7 shows how to assign an allowance for the virtual probability "in transit" along a light cone, so that the calculation of the total probability on any spacelike surface that slices across the spacetime diagram is conserved.

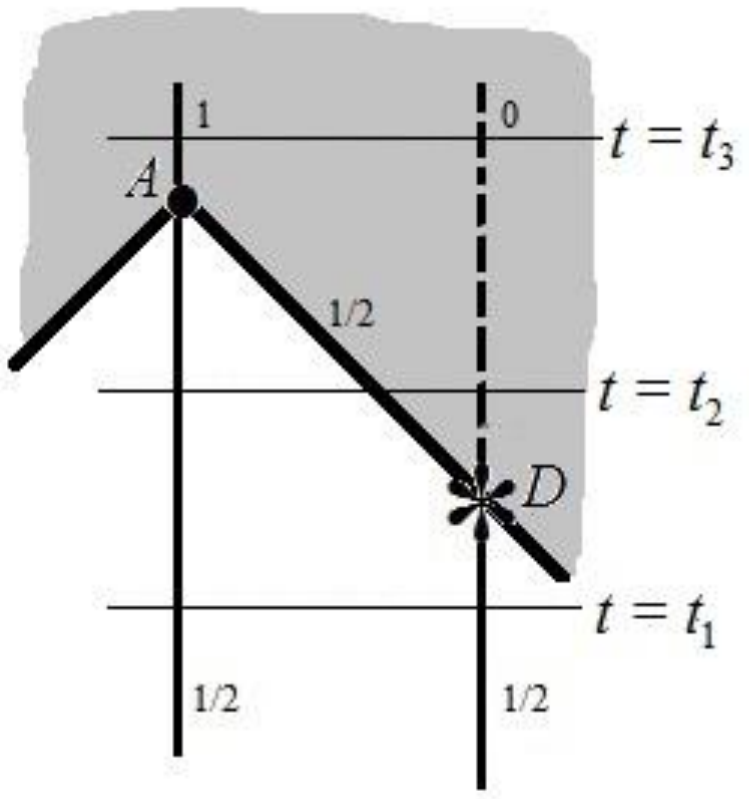

**Fig. 7** Probability weights for the two timelike worldlines of two cards and for the lightlike segment *DA* of the past light cone of the inspection point *A* . For all the spacelike surfaces (thin horizontal lines) cutting across this diagram the total sum of probability weights is the same.

The lesson learned this example of collapse of a spatial probability distribution of playing cards is that the requirement of Lorentz invariance of probabilities (that is, the behavior of probability as a Lorentz scalar) in conjunction with probability conservation inevitably leads to the conclusion that spatial probability distributions collapse along the past light cone of the measurement point. If measurements are performed at several points, then collapses occur along the several corresponding past light cones. The net result is that the collapse extends over the earliest portion of the surface of the union (topological $\bigcup$) of all the several light cones. And this is, of course, also true for measurements at a continuum of measurement points spread over a continuous spacelike hypersurface.

Unfortunately, in our Queen-of-Spades game it is not possible to perform a direct empirical test of the pre-collapse phenomenon of the relativistic past-light-cone scenario. No inspection of a card can affect the underlying physical state of the probability distribution or affect the outcome of any other inspection. The pre-collapse is merely a deduction from probability conservation, which permits us to learn something about the probability distribution earlier than by direct inspection; that is, the pre-collapse merely affects what we know when and where, but it does not affect the ultimate outcome of the game—who wins and who loses—which is predetermined by the initial conditions imposed by the initial mailing of the playing cards. In contrast, in a quantum-mechanical probability distribution, inspections (or measurements) of wavefunctions do affect the wavefunctions, and in particular the coherence of the wavefunctions. In Section VI we will see how to exploit this for an empirical test of pre-collapse.

Does the past-light-cone scenario have any practical consequences at all for playing the Queen-of-Spades game? If each player is honest and does not take unfair advantage of "insider" information about how many cards have already collapsed to zero—by, say, offering to buy a sealed card from another player who does not yet know how many cards have already collapsed to zero—then there are no practical consequences. In the game of chance, the location of the Queen is unknown to the players but is fixed, and the inspections do not alter the physical location of the Queen.

However, an understanding of relativistic collapse is helpful if we want to prevent unfair "insider trading," and we want to prescribe a schedule of card inspections that eliminates such unfair trading. Obviously, one way to do this is to schedule the card inspections worldwide in synchrony at, say, Universal Coordinated Time 10:00:00. The relativistic collapse scenario tells us there are many other schedules for card inspections that are equally good at preventing unfair trading. Any schedule that places the spacetime points of card inspections on some spacelike hypersurface, flat or curved, that passes through all these points is an allowed schedule. A limiting case for such allowed schedules is a hypersurface that asymptotically approaches the past light cone of one of the inspection points. Such a collapse along the past light cone of an inspection point, with all other inspection points along this same light cone, is the simplest way to eliminate insider trading. This establishes a financial connection between past-light-cone collapse and fair trading practices—and money talks, doesn't it?

Finally, a comment about collapse on a cosmological scale. The generalization of the collapse paradigm for the curved spacetime of Einstein's relativistic theory of gravitation seems obvious: instead of the rotationally symmetric light cones of special relativity, we must rely on the deformed light cones that result from the "bending" of light rays by gravitational effects. For a sufficiently large gravitating mass near the path of the light rays, this bending can result in multiple images of a pointlike source, which implies folds in the future light cone. Correspondingly, the past light cone of a spacetime point will have similar folds, or overlapping sheets, in a sector lying beyond the place where the past light cone intercepts the gravitational lens. In this case, the correct choice for the surface of collapse for the probability distribution will have to be the innermost sheet of the folded light cone. The full implications of such folds are not immediately obvious.

## IV. Collapse of quantum-mechanical wavefunctions

We will now apply the lessons learned about relativistic collapse of probabilities in a game of chance to the relativistic collapse problem in quantum mechanics. Of course, there is an important difference between the probability collapses in our game of cards and that in a quantum-mechanical system. In our game of chance, the collapses triggered by the several measurements, or inspections, of the cards do not alter the physical state of the cards, which is predetermined once the cards have been inserted in their envelopes and sent off to their destinations. The probability distribution of cards is merely a measure of our ignorance about the true physical state, and the collapses in the distribution are changes in the information we have available about this true physical state. In contrast, in a quantum-mechanical system the collapses alter the physical state. As DeWitt expressed it, in quantum mechanics, "chance is not a measure of our ignorance but is *absolute*." [32]

Besides, in a quantum-mechanical state vector or wavefunction[33] distributed over several spacetime locations, the several parts are coherent, with well-defined phases for all the parts, whereas for a distribution of playing cards there is no such coherence. But this quantum coherence terminates just before the final collapse of the quantum-mechanical probability distribution, because the measurement proceeds in three steps:[34] (*i*) First the system interacts with an apparatus by normal quantum-mechanical interactions between the system and the quantum-mechanical entities in the apparatus (photons, electrons, atoms, nuclei and nucleons, etc.). This leads to a coherent superposition of entangled states of the system and the entities in the apparatus, with amplitudes that correspond to the Born rule. (*ii*) Then decoherence comes into play and converts the coherent superposition into an incoherent superposition (or a "mixture"). Such an incoherent superposition is a probability distribution, analogous to the probability distribution in a game of chance, and this permits us to treat the subsequent behavior in the same way as in a game of chance. (*iii*) Finally, according to the Copenhagen interpretation, the intervention of an external observer triggers the mysterious "arbitrary" von Neumann Process 1 that collapses the probability distribution (or, according to Everett, the observer "perceives," and the world splits into several branches; or, according to Griffiths, the observer elects to end one epoch of history and begin a new epoch; or, according to your favorite *chef de cuisine*, something else happens—*chacun a son gout*). In this final collapse (or splitting, or historical accounting, or something else) we can apply what we have learned about past-light-cone collapse in a game of chance.

For the terminal collapse of the probability distribution in the third step, the past-light-cone collapse scenario is merely a kinematical algorithm that makes the collapse consistent with the geometry of relativistic spacetime and Lorentz invariance. It does not provide any dynamical foundation for the collapse process, which is to be regarded as a separate though not entirely independent issue—any proposal for a relativistic dynamical mechanism of wavefunction collapse that does not obey the relativistic kinematics of the past-light-cone collapse scenario falls under vehement suspicion of a logical flaw.

In all the published discussions of quantum measurements, the relativistic collapse of quantum-mechanical probability distributions was examined without the helpful guidance offered by the simpler case of a game of chance. This has tended to leave the fundamental spacetime aspects of the collapse problem murkily enmeshed with quantum-mechanical metaphysics, which perhaps accounts for much of the confusion found in discussions of relativistic collapse

Quantum-mechanical probability distributions are usually continuous distributions and, furthermore, measurements often span a region extended over a spatial line, or a surface, or a volume, and also span a time interval, for instance the volume of a multi-wire particle detector used in a high-energy accelerator experiment that a particle transverses in a finite time interval. The collapse generated by such a continuum of measurement points is then a generalization of the collapse for a discrete set of measurement points illustrated in Fig. 5. The resulting uncollapsed region is the topological intersection of all the absolute past regions of all the light cones associated with the continuum of measurement points; and the fully collapsed region is the complement of the closure of the union of all the absolute past regions. These regions are shown in Fig. 8. In the intermediate transitional region between the collapsed and uncollapsed regions the collapse is in progress—it has started, but has not yet been completed. Note that the collapse starts before the measurement begins; if the size of the measurement region is $d$, the collapse typically begins at a time of about $d/c$ before the measurement. In Section VI we will see how it might be possible to take advantage of this pre-collapse phenomenon to test past-light-cone collapse.

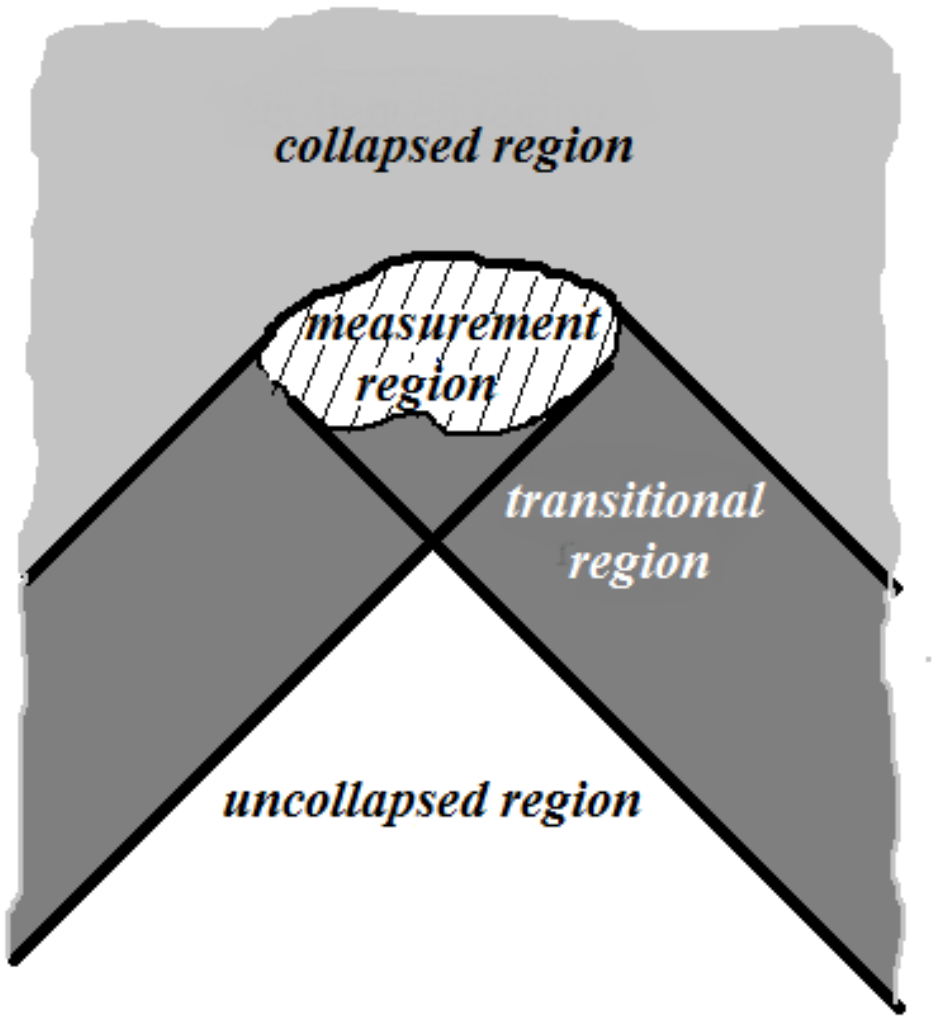

**Fig. 8** Collapse of a probability distribution along the multitude of light cones associated with a continuous distribution of measurement points extended over a finite spacetime region (shown hatched). The fully collapsed region (gray, at top) is separated from the uncollapsed region (white, at bottom) by a transitional region (dark gray).

The possibility of collapse along the past light cone in quantum mechanics was first explored in 1967 by Bloch, who recognized the advantage of adopting a Lorentz-invariant hypersurface—such as the past or future light cone or an invariant hyperboloid—to attain Lorentz invariance of the collapse process. However, Bloch rejected this on the grounds that the quantum-mechanical wavefunction "would be noncausal, and could not possibly be calculated from a differential equation with initial conditions, " because "at all times before $t_a$ [or $O$ in my notation] the function would already be influenced by the result of the observation…which is not made until the time $t_a$." And he claimed that either causality or Lorentz invariance of the wavefunction must be sacrificed.

But this argument is wrong: there is no reason why the collapse needs to result from the operation of a differential equation. We are dealing with the collapse of a probability distribution, which, as in a game of chance, is not compelled by a physical process with causal propagation of wavefronts or long-range interactions, but by the logical imperative of probability conservation, which tells us that if a particle or some other position-dependent system state is found at one location, then the probability weights at other locations must collapse, without the need of any propagation of anything physical between these locations (we can of course exploit physical propagation of signals as an adjunct for investigating the collapse, but we are not obligated to do so).

In their 1981 and 1984 papers,[25] Aharonov and Albert re-examined the Hellwig-Kraus collapse along the past light cone and conceded that it yields the correct probabilities for local measurements, but they strenuously argued that for nonlocal measurements, with linked measurement devices extending over a spatial region, collapse along the past light cone yields inconsistent results. They gave a long, tedious counterexample to past-light-cone collapse based on measurement of the joint spin of two particles separated by some distance. They argued that in this case it is possible to contrive a “nondemolitional” measurement procedure for the total spin without measuring the individual spins. Their procedure used two linked spin-detection devices located at the positions of the particles and arranged in a carefully tuned initial state, for which they demonstrated that the sum of the outputs of the devices gives a measurement of the total spin while leaving the individual spins unmeasured.

The argument of Aharonov and Albert was quashed by Mould,[35] who showed that they had neglected an essential extra step in their measurement procedure: the need to bring the outputs from the separate measurements of the spin detectors together, so as to generate a net result for the total spin, which could then be subjected to collapse by von Neumann’s Process 1. In essence, Mould showed that to bring the measurement to a final completion we need to combine the outputs of two the spin detectors by physical links, such as electric wiring, or optical fibers, or other physical transmission devices that carry these outputs to some single location, say, the midpoint between the two detectors, where a macroscopic “observer” can perform the necessary algebra on the outputs by analog or digital means and thereby complete the experiment. If so, then the trigger point for the collapse of the probability distribution for the various possible outcomes of the spin measurement is this “observer” at the midpoint—which means we have a simple case of collapse triggered by measurement at a single spacetime point. The collapse then proceeds along the past light of this spacetime point, as in Fig. 3.

The instantaneous $t = 0$ hypersurface that Aharonov and Albert use to span the interval between their two detector devices is merely a figment of their imagination. Their experimental apparatus does not perform any operations in that interval, that is, their experimental apparatus would operate equally well if all the spacetime points within that interval were blocked out and declared a forbidden zone. The only time parameters that Aharonov and Albert use in their calculation are the times along the worldlines of the particles, for which they could just as well use the proper times. Their adoption of the instantaneous $t = 0$ hypersurface to define a collapse time is a possible but not a compelling choice. Their attempt to refute the past-light-cone collapse with their example of a nonlocal measurement of a pair of spins therefore fails, because this refutation rests entirely on a false claim of a compelling need for instantaneous collapse in nonlocal measurements.[36]

Any nonlocal measurement consists either of several independent measurements distributed over a set of spacetime points—in which case we can treat it as set of individual collapses, so the hypersurface for the start of collapse is the surface of the union of the individual past light cones (as in Figs. 5 and 8). Or else it involves a sum (or some other algebraic combination) of outputs from several measurement devices placed at a set of spacetime points, and the experimental procedure for the summation (or other algebraic combination) of these outputs then requires transmission links that carry the outputs to one location where the numerical combination is performed by a counting or metering device—in which case we can regard the location of the counter as the collapse point, so the hypersurface of collapse is the past light cone of the counter. When and where a state vector collapses is decided by the experimental equipment and how it is operated, not by the dreams and fancies of theoreticians.

Furthermore, Aharonov and Albert suffered from the misconception that it is permissible to extend the spatial measurement region across a past-light-cone boundary, so the spacelike hypersurface on which this measurement is performed lies partially in one layer of world and partially in the next (see $t = t_2$ in Fig. 4). This is just as daft as to propose an instantaneous nonrelativistic collapse scenario defined by the Newtonian absolute-time hypersurface $t = 0$, and proceed to perform a time-nonlocal measurement (say, a time-average measurement) that spans a time interval from $t = -1$ minute to +1 minute, using a mindless mix of probabilities from before and after the collapse. Aharonov and Albert relied on an example containing such forbidden boundary-crossing measurement regions to argue that the monitoring of a system by a sequence of nondemolitional measurements in one reference frame is inconsistent with such monitoring in another reference frame. But their argument merely tells us if we want to monitor a system we must make sure that any nonlocal measurement, performed in whichever reference frame we choose, lies entirely within one and only one layer of the spacetime "layer cake" (see $t = t_1$ in Fig. 4).

Thus, all the objections made by Aharonov and Albert in their examination of nonlocal measurements are untenable. This is not to say that nonlocal measurements extending over regions of space or time are free of complications; but whatever these complications are, they are not the result of adopting the past-light-cone collapse scenario.

The objection made by Cohen and Hiley[37] to the pre-collapse phenomenon illustrated in Fig. 6 is also untenable. It amounts to no more than an obstinate assertion of their belief that such a collapse that happens before "either particle has been subjected to a measurement…seems absurd." Cohen and Hiley fail to grasp that when we say that the collapse happens "before," this merely means it is *nominally* before, that is, it is before in some references frames but it is not before in all reference frames. The point *D* in Fig. 6 lies in the ambiguous region relative to the point *A*. This ambiguous region is an open set, and we can always find a high-speed Lorentz frame whose equal-time hypersurface $t' = 0$ is nearer to the light cone than the point *D* at which the collapse occurs (as already emphasized in Section III, the collapse is not *on* the light cone, but *along* the light cone, that is, just outside of the light cone). Thus, the usual ambiguity of time order for the ambiguous region applies: *D* can be nominally before or after *A*, depending of what reference frame we adopt. Maybe the reversal of (nominal) time order between *A* and *D* is puzzling (for people not accustomed to the relativity of synchronization), but it is not absurd or illogical. The point *D* is not in the absolute past of *A*, so there is no reversal of the absolute time order, and there is no question of a paradoxical causal loop.

The collapses of the probability distribution are not physical processes brought about by interactions or signals (which can act only in the absolute future), but logically compelled changes imposed by the imperative of probability conservation. These logically compelled changes are causal in a general sense of causality, in that *D* is implied by *A*, even though the nominal time order is reversed in the *t*-*x* reference frame implicit in Fig. 6. In Newtonian spacetime, we are well-accustomed to the instantaneous collapse of probability distributions in a game of chance. The past light cone in Fig. 3 merely takes over the role of the instantaneous $t = 0$ collapse hypersurface of Newtonian spacetime by demanding that it must hold invariantly in all inertial reference frames, and the pre-collapse phenomenon is then a necessary consequence. Perhaps this is weird, but it is not absurd, that is, against logic. And, of course it is no more weird than many other phenomena we have become accustomed to in the realms of relativity and quanta.

## V. Collapse on two beamlines and collapse in EPR experiments

An important special case of collapse involves two coherent narrow wavepackets traveling along two separate beamlines. The beamlines might be parallel, as in the case of an atom split into two wavepackets traveling along parallel paths in an atom interferometer; or they might be divergent or convergent after reflections at the atomic gratings in the interferometer. In the EPR Gedankenexperiment, whether with atoms or photons, the beamlines are divergent. For nonintersecting portions of the paths, the topology of collapses produced by measurements of the positions of the atoms are essentially the same as in the example of two playing cards on parallel worldlines discussed in Section III.

Figure 6 can be regarded as describing the situation in a reference frame in which the atom wavepackets are rest (the rest frame of the atom). Detection or nondetection of the atom can occur either at a spacetime point *A*, or *B*, or at both *A* and *B*. Each of these cases leads to a different collapse hypersurface, consisting either of the past light cone of *A*, or of *B*, or of both *A* and *B*. For detectors operating at both *A* and *B*, one of these detectors could be considered redundant because upon detection of the atom at, say, *A* we know it is not at *B*.

In the nonrelativistic theory of collapse of quantum measurements, an attempted detection at *B* with a confirmatory null result is called a nondemolitional measurement, because it is regarded as a mere re-measurement of something we already have established previously, and it is generally assumed that such a confirmatory re-measurement does not disturb the quantum-mechanical state at all. However, in our relativistic theory of collapse we see that the seemingly repetitive detection measurement at *B* does change the collapse boundary: instead of the single past-light-cone boundary of Fig. 6(a), we now have the two past-light-cone boundaries of Fig. 6(c), which jointly lead to a smaller uncollapsed region (white) and also a smaller final collapse region (gray) than Fig. 6(a). Thus, "nondemolitional" measurements do produce significant changes in the boundaries of collapse. But, of course, these changes become noticeable only if the spatial extent of the system is large enough so the time delays *CA* and *DB* cannot be treated as negligible.

Past-light-cone collapse also gives us a neat description of the EPR experiment and other similar experiments involving measurements on a coherent state vector or wavefunction at two locations separated by a distance sufficiently large to exclude communication within the time interval allowed for the measurement. Instead of collapse of the wavefunction of a single particle distributed over two beamlines, as in Fig. 6, the EPR experiment uses two particles with entangled wavefunctions or spin states. For the sake of simplicity, discussions of the experiment often use two electrons in an entangled spin state, sent away in opposite directions where they are intercepted by detectors that measure their spins. If the net spin of this two-particle system is zero, and one of the detectors finds the intercepted electron to be in a spin-up state, then the other electron must collapse into a spin-down state.

Relativistically, this collapse must proceed along the past light cone of each detection point, so the general features of the collapse are similar the two-wavepacket case discussed in the preceding paragraphs, even though we are now dealing with two individual particles, not two wavepackets belonging to a single particle. The other change in the collapse diagram is that the two worldlines originate at a common point *P*, where their two spins were assembled into the entangled spin state. Figure 9 shows these two worldlines for two particles that move away from each other in opposite directions along the *x* axis, and it shows the past-light-cones for the collapses triggered by spin measurements at the two detectors *A* and *B*.

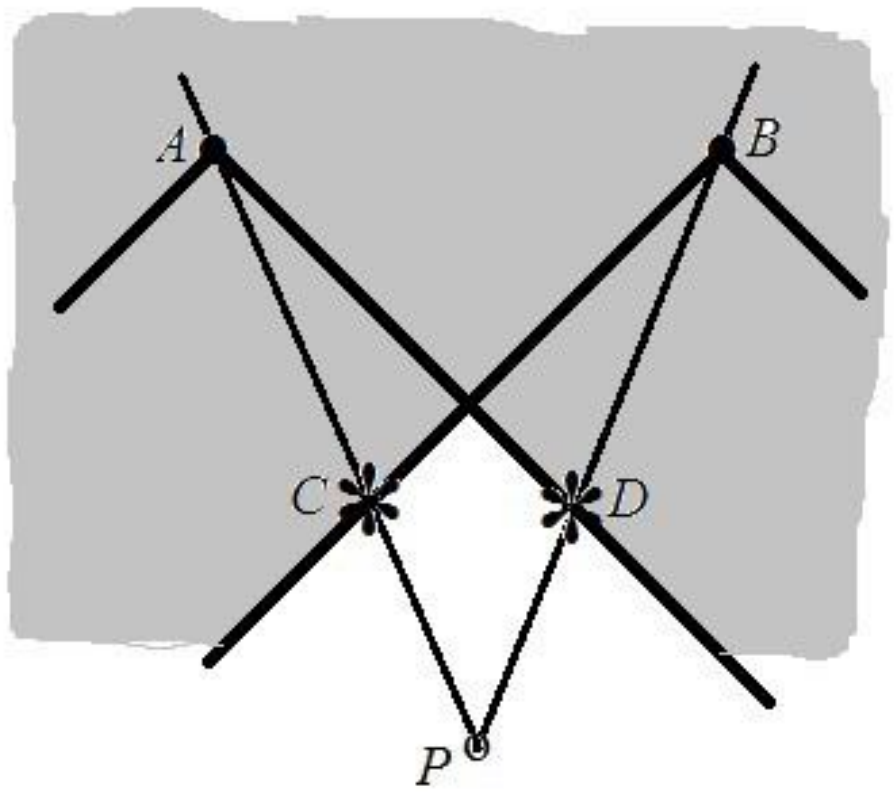

**Fig. 9** Worldlines of two particles traveling away from the point *P* in opposite directions. Their entangled spins are detected at the points *A* and *B* and their wavefunctions and spin components collapse along the past light cones of *A* and *B*.

If the two particles move at very high speed ( $v \to 1$ ) their worldlines actually lie very near the past light cones, and the collapses along the past light cones of *A* and *B* extend almost all the way to the point of origin *P* of the particles. If we perform such an experiment with photons, this backward collapse extends *exactly* to the origin, that is, the photons collapse into their final state immediately after the instant at which they are placed in their initial state (see Fig. 10). For a photon, the final state “happens” at the next instant after the initial state, as we might intuitively expect because of the zero proper time interval ( $d\tau = 0$ ) that characterizes light-like worldlines.

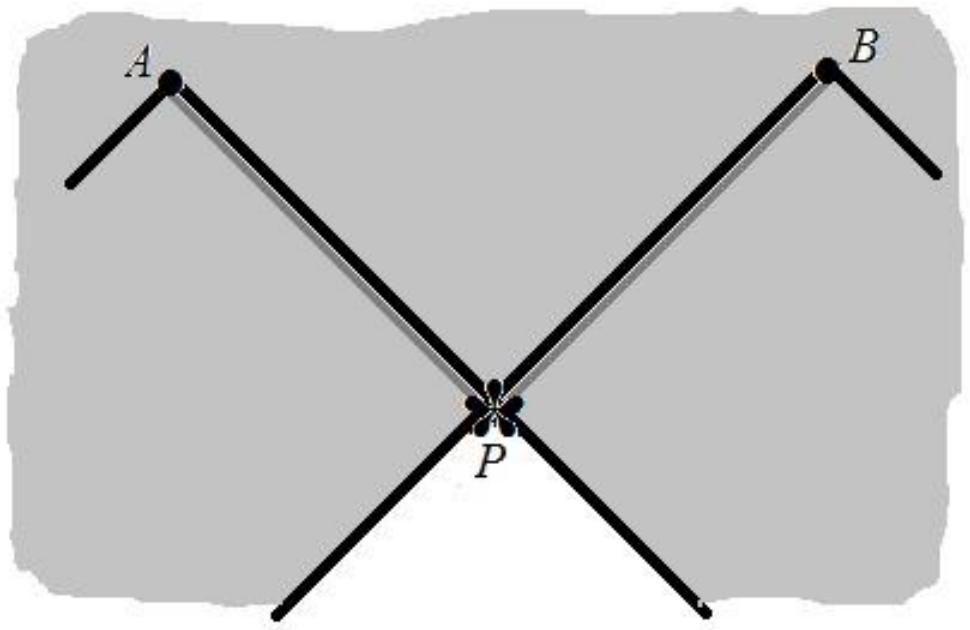

**Fig. 10** Worldlines of two photons traveling away from the point $P$ in opposite directions. The photon worldlines (gray) are shown displaced slightly downward, to distinguish them from the segments of past light cones (black) that overlap them.

Such a surprising collapse near the point of origin of the photons is also applicable to photons travelling on parallel beams, for instance, two photons with a constant transverse separation *d* in the *x* direction while they travel in the *z* direction, so the worldline of one photon lies in the *x* = 0 plane (that is, the *t*-*z* plane) and that of the other in the *x* = *d* plane. By examining the past light cones of the detection points *A* and *B*, we can recognize that each of these past light cones fails to intercept the worldline of the other photon except in an early region near the source (say, a laser), where the two photons were diverging from a common point and their transverse distance in the *x* direction was increasing.

This has a direct consequence for tests of hidden-variable theories, such as the test performed by Aspect et al.[38] with two entangled photons sent from a common source to two detectors at opposite ends of the laboratory. The collapse in this case does not occur at the two distant detectors, but right at the source, so it is quite possible that the outputs of the detectors influence each other—which invalidates the claim that the experiment is a test of instantaneous action-at-a-distance. To avoid this problem, it is necessary to send the photons to their destination by a roundabout route or to place the photons in a holding pattern before the detection, which deforms their worldlines into short zig-zags before they reach the detectors. This delays the collapses for some fairly long time, so the past light cones of the measurement points pass well above the point *P*. (This condition was satisfied in an experiment by Tittel et al.[39] who used optical fibers of lengths of 8.1 km and 9.3 km, respectively, for sending the photons to the two detectors, which were separated by a straight-line distance of only 10.9 km instead of 8.1 km + 9.3 km. But the use of optical fibers might be problematic, because in a refractive medium the photons scatter, and this by itself might be regarded as a measurement?)

Apart from this correction in the experimental procedure, tests of local hidden-variable theories via Bell’s theorem remain valid, because the essential ingredient in this theorem is that for such hidden variables the initial state of the system is laid down and held fixed at the source, so the probability distribution merely represents our ignorance of this distribution; whereas for a quantum-mechanical system the state is altered by each of the measurements that are performed during the experiment.

## VI. Experimental tests of past-light-cone collapse?

It seems desirable to contrive some experimental test that confirms or refutes past-light-cone collapse and, in particular, the pre-collapse phenomenon. As already mentioned in Section III, in a game of chance such a test is impossible because the underlying physical state of the probability distribution is permanently fixed by the initial conditions and is not altered by measurements (that is, by inspections), so the pre-collapse merely means that we can deduce something from probability conservation earlier than we can confirm it by direct inspection. However, for a quantum-mechanical probability distribution, each measurement changes the underlying physical state in both its (absolute) amplitudes and its phases, that is, in both its probabilistic outcomes and its phase relations. By a sufficiently gentle measurement, we might be able to detect the pre-collapse decoherence of the collapsed state without disturbing the probabilistic outcomes, which are constrained by probability conservation. As we will see, an examination of the relative phases of two coherent wavepackets in the beams of an interferometer offers an opportunity for a test of pre-collapse and therefore a confirmation of past-light-cone collapse.

In a previous publication,[40] I proposed how such an experimental test might accomplished by atom-interferometry, in imitation of the techniques used in the ingenious atom-interferometer experiment of Chapman et al.,[41] which was designed to explore how two coherent atom beams respond to probing by a transverse laser beam whose photons reveal, or partially reveal, the positions of the atoms and thereby introduce decoherence, or partial decoherence, between the atom beams. Here I will add some more details to this proposed atom-interferometer test, and also propose an alternative test that relies on an ordinary light interferometer.

In the Chapman et al. experiment, a beam of sodium atoms was sent into a Mach-Zehnder interferometer, where diffraction gratings split the atom beam into two spatially separated coherent beams, with a transverse separation of about 2 mm. The two atom beams were probed with a focused laser beam aimed across their paths, so the laser light was scattered by fluorescence, that is, photons were absorbed by the coherent atom wavepackets while crossing the focal spot of the laser beam and then promptly re-emitted by spontaneous fluorescent decay (with an emission half life of about 17 ns). The two coherent atom beams were then recombined in the interferometer to allow them to form interference fringes which were scanned with a hot-wire atom detector placed in the recombination region.[42]

Chapman et al. found that if in their measurements they left the direction of scattering of the photons completely undetermined, the decoherence was large, and the interference pattern generated by the recombination of the beams in the interferometer lost its contrast. They also found that if their measurements selected a limited range of angles of emission for the photons, the decoherence was small (or, as they expressed it, “coherence is regained”). This is as expected, because if the final state of the photons is constrained to a known direction, then radiation of the photon introduces no uncertainties into the motion of the atom wavepackets and does not spoil their coherence.

The experiment I propose would use much the same arrangement as that of Chapman et al. for generating two parallel coherent atom beams and for probing these with a transverse laser beam. But beyond that stage the experiment would rely on examination of the radiation pattern of the fluorescent laser light and on atom detectors at the end of each beam line to trigger a final collapse of the atom at the detection points *A* or *B* (see Fig. 11).

If these final location measurements trigger pre-collapses along the past light cones of the detection points *A* and *B*, the atom beams will become decoherent at *C* and *D*, and the fluorescent radiation emitted over the intervals *CA* and *DB* will be that of two incoherent antennas. If the final location measurements do not trigger pre-collapse, the fluorescent radiation emitted over the intervals *CA* and *DB* will be that of two coherent antennas, with a phase difference that depends on the distance between the atom beams. Beyond the detection points *A* and *B*, the fluorescent radiation would be incoherent, but because the atomic states will probably be mangled by the collisional interaction of the atom with the detectors, there will probably be no fluorescent radiation at all.

In the region *AWBPA*, the radiations arriving from the left and right wavepackets of each atom overlap and, if coherent, the superposition of the radiations arriving from the left and right worldlines will form a standing e.m. wave, with maxima at intervals of half a wavelength. These maxima can be detected by a strip of photographic film placed along the midline between the two atom beams, so the standing e.m. wave reveals its presence by imprinting a fringe pattern in this film (similar to the fringes that are imprinted in a fine-grained photographic emulsion in the production of a hologram). Each pair of wavepackets for each atom that passes through the interferometer contributes the same intensity pattern of fluorescent light waves at the location of the film strip, so adequate statistical data to confirm or refute the existence of interference fringes can be accumulated by passing a large number of atoms through the apparatus. If the experiment reveals interference fringes, it would refute the past-light-cone-collapse scenario.

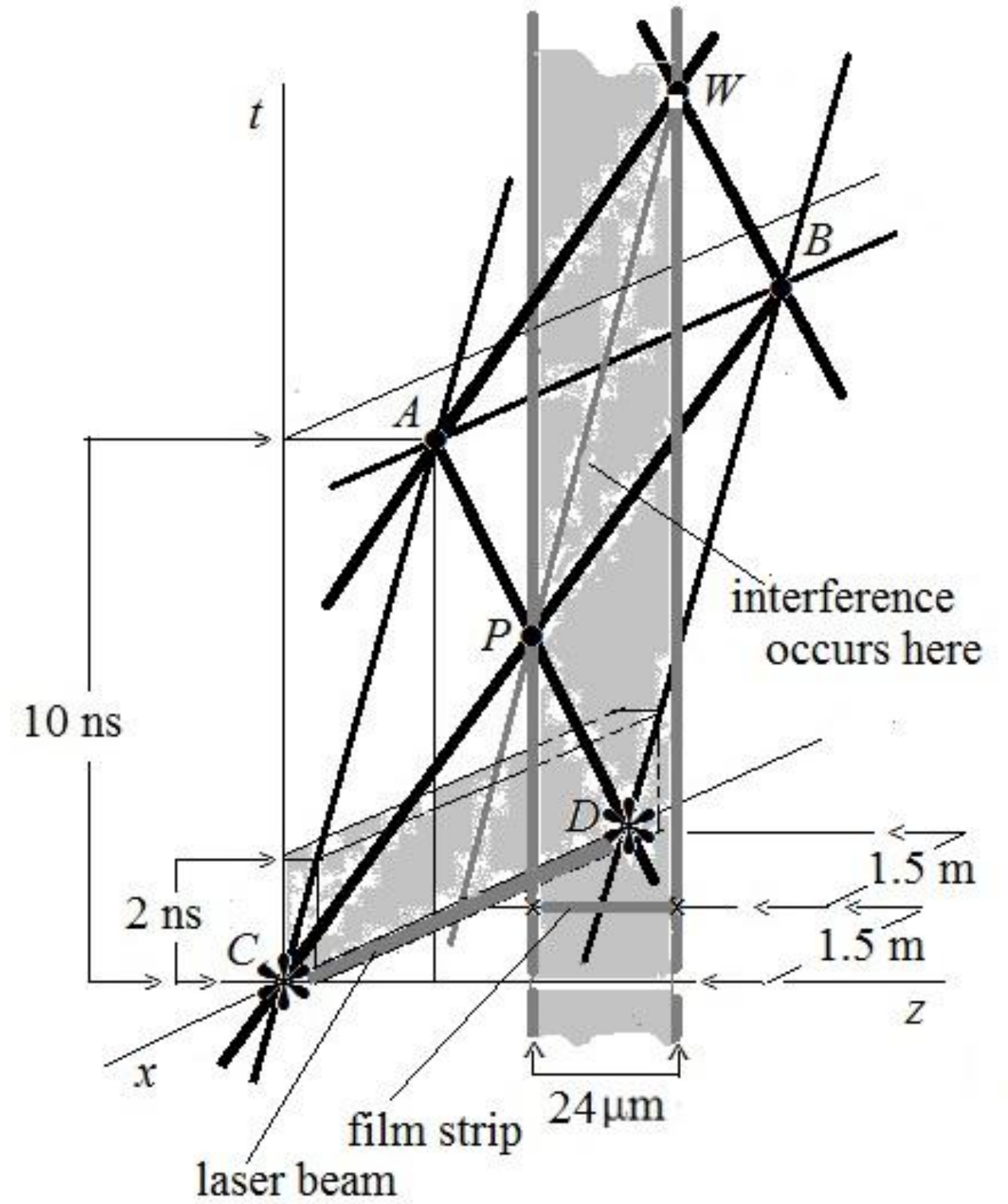

**Fig. 11** In this spacetime diagram, *CA* and *DB* are the worldlines in the laboratory frame *t*, x, *z* of the two wavepackets of a single coherent sodium atom traveling at 3000 m/s along the parallel beamlines in an atom interferometer (the wavepackets travel in the *z* direction; the slant of their worldlines relative to the *t* axis has been exaggerated by a factor of about $10^5$ for clarity). Hot-wire detectors at *A* and *B* collapse the wavepackets and determine whether the atom is located at the left detector or the right detector. The diagram assumes a transverse distance (in the *x* direction) of 3 m between the beamlines, and consequently a pre-collapse time of 10 ns, which allows enough time for absorption of a photon from the laser beam and subsequent emission of a fluorescent photon. Just after reaching the pre-collapse points *C* and *D*, the wavepackets cross the laser beam (at rest in the lab frame) in which they spend 2 ns and absorb a photon (the spacetime region, or worldtube, of this laser beam is indicated by the rectangular box; only a 2 ns segment of this worldtube is shown here, which is where the wordlines cross the worldtube). The 8 ns that elapse after the emergence of the worldlines from the laser beam and arrival at *A* or *B* is about ½ of the half-life for spontaneous fluorescent emission, so about 30% of the excited atoms will radiate before arrival at *A* and *B*.

If the naïve scenario of instantaneous collapse on a $t$ = constant hypersurface between *A* and *B* is valid, the emissions from the two wavepackets proceed coherently, and the photon state at or near the midline between the wavepacket wordlines is a standing wave, with nodal points of constructive interference spaced half a wavelength apart in the *x* direction. This constructive interference lasts from the spacetime point *P* to *W*, and beyond *W* further superposition of radiation from remaining excited atom wavepackets will be incoherent. However, before the coherent interference ends, the enhanced photon concentration at the nodal points will have made a detectable imprint on a small strip of photographic film (a "microdot," about 24 $\mu$m long and a few $\mu$m wide) placed at rest in the laboratory frame at the midpoint between the detectors *A* and *B* (the worldtube of this microdot is shown as vertical gray band in the diagram). If the past-light-cone collapse scenario is valid, emissions from the two wavepackets proceed incoherently for the segments of worldlines beyond the pre-collapse points *C* and *D*, and there will be no nodal points of constructive interference of the fluorescent light.

The device proposed for this experiment could be called a hybrid atom-light-interferometer: it is a combination of the front half of an atom Mach-Zehnder interferometer and the back half of a light Mach-Zehnder interferometer, and it operates by transferring the phase information about coherence (or incoherence) of the paired atom wavepackets to the light that will be intercepted by the film strip. Note that, because the film strip is at the center between the beam lines and has no directional sensitivity, the capture of a photon by this film strip conveys no "welcher Weg" information, that is, it does not tell us whether the atom is on the left or on the right. This position information is acquired only later, by the atom detectors at *A* and *B*. Thus, capture of photons by the film strip does collapse the atom wavepackets and does not preempt the position detection at *A* and *B*.

The size of the atom interferometer apparatus is determined by the requirement that the time interval *DA* must accommodate the time for the laser excitation of the atom (which is assumed to be 2 ns, attainable with a reasonable laser power density[43]) followed by spontaneous fluorescent decay of the excited state (the decay interval of 8 ns assumed for Fig. 11 is about ½ of the half-life of the excited state, which results in decay of 30% of the excited atoms; the remainder of these atoms either decay incoherently, without contributing to the enhancement at the maxima, or they do not contribute at all because at the detectors at *A* and *B* they experience collisional disturbances that abort the emission process. Thus, the 3-m distance between the beams is determined by the need to make enough time available for the excitation of the atoms followed by decay of a reasonable fraction of the excited atoms. In the Chapman et al. experiment, the distance between the beams was about 2 mm, so an "upgrade" of the atom interferometer by a factor of somewhat more than 1,000 is needed, if the experiment is to be performed with sodium atoms.

Because of the difficulties in constructing such a large atom interferometer, it might be preferable to conduct the experiment with a light interferometer, for which a size of 3 m would not be problematic. Suppose that the two coherent parallel atom beams in Fig. 11 are replaced by two coherent light beams, generated by splitting a Ne laser beam ( $\lambda = 632.8$ nm) with a grating or a half-silvered mirror. Then, by past-light-cone collapse, the detections of individual photons by detectors at *A* and *B* trigger pre-collapses at *C* and *D*, and the light-beam segments *CA* and *DB* become incoherent. We can detect this incoherence by inserting He-Ne discharge tubes into the paths of the light beams, beyond *C* and *D* (instead of the laser beam shown in Fig. 11). In these discharge tubes, the collisions between Ne and He atoms optically pump the Ne atoms to their excited state, and the light beams passing through the discharge tubes then generate radiation of extra Ne light by stimulated emission (in the absence of a resonant chamber terminated by reflecting mirrors, the stimulated emissions from the He-Ne discharge tubes will merely take the form of omnidirectional dipole radiation, instead of unidirectional laser beams).

If the light beams *CA* and *DB* are incoherent, the stimulated emissions they create during transit through the discharge tubes will also be incoherent. If the light beams are coherent, the stimulated emissions they create during their transit will be coherent in the forward direction of the beam lines (as in "forward scattering," in the *z* direction), but incoherent in the transverse directions, because the emissions have different points of origin at randomly placed Ne atoms in the discharge tubes. However, for emission directions very close to the beam lines (say, within 0.1 milliradian), the phase errors accumulated over the width of the discharge tubes are small, and the stimulated emissions remain approximately coherent. This means that we can "skim off" coherent stimulated emissions by means of deflecting mirrors placed very near the left and right beam lines, and we can deflect these coherent stimulated emissions toward the region between the beam lines, where the emissions will interfere and thereby reveal their coherence. Rough estimates of the parameters needed for the operation of such a light-interferometer indicate that the experiment might be within practical reach.

Thus, in both of these proposed experimental tests of past-light-cone collapse, this collapse is confirmed or refuted by absence or presence, respectively, of coherent fluorescent or stimulated emissions.

## VII. Conclusions and Some Further Comments

The relativistic treatment of the collapse of probability distributions in a game of chance presented in the early sections of this paper establishes the necessity for collapse along the past light cone. For the later sections of this paper, this lays a firm foundation for a similar collapse of quantum-mechanical probability distributions and wavefunctions. However, the collapse along the past light cone merely provides a description of the relativistic kinematics of quantum-mechanical collapse—it does not explain why one of the various probabilistic alternatives is selected and the others are rejected. In a game of chance, this selection process is performed by the player or the "dealer" of the card game, who selects an unknown card by his or her free will. But in quantum mechanics we have no such obvious selection mechanism, and the relativistic kinematics of collapse presented here do not offer any enlightenment about this fundamental problem of quantum-mechanical measurement theory.

Collapse along the past light cone makes the quantum-mechanical collapse process seem even more mysterious than the naïve nonrelativistic collapse at one instant of time, because in most reference frames the collapse begins in the distant past at the furthest boundaries of the system and proceeds inward at the speed of light, to end at the measurement point. This is spooky action-at-a-distance with a vengeance—not only does the collapse extend over long distances of space, but it also extends backward in time. However, from the discussion in Section II , we know that we can always find a high-speed reference frame in which this time order is reversed, so the measurement point precedes any given collapse point alongside the light cone, a curious circumstance that hinges on the collapsed region being an open set (this subtle detail in the topology of the collapsed region will no doubt confuse and awe students in introductory quantum mechanics).

An important corollary of the general treatment of the relativistic kinematics of collapse presented in this paper is that past-light-cone collapse must also be applied to interpretations of quantum mechanics other than the Copenhagen interpretation. Thus, in both the Everett many-worlds interpretation and the Griffiths consistent histories interpretation the hypersurfaces of discontinuity, on which the world splits into several branches (Everett) or on which we elect to end one epoch of history and begin the next (Griffiths), cannot be flat spacelike hypersurfaces—they must be past light cones.

Such a replacement of flat hypersurfaces by light cones is easy to implement if the successive light cones do not intersect (as in Fig. 4 ), so we can assign a well-defined time order to these cones, and each cone lies entirely within the absolute past of the next cone. But it becomes problematic if the cones intersect (as in Fig. 5), and the apex of each cone lies outside of the absolute past of the other cones, so the time order of the apexes of any two intersecting cones can be reversed by a Lorentz transformation, and the absolute past of none of the cones lies entirely within the absolute past of another. This implies that we cannot construct an unambiguous progressive history with epochs bounded by these cones. For instance, a history that begins on the $P_1$ cone, progresses to the $P_2$ cone, and ends on the $P_3$ cone in Fig. 5 is possible, but so is a history that progresses from the $P_3$ cone to $P_2$, and then to $P_1$. For the consistent-histories interpretation this means that not only do we need to impose on these histories the usual requirement of consistent probabilities, but we also need to impose a requirement of nonintersection of the light cones that we use as boundary hypersurfaces separating one epoch of the history from the next.

In the Everett interpretation the past light cone of a measurement point is not a hypersurfaces of collapse, but a hypesurface of branching on which there is a transition from the initial single world to an ensemble of many branched worlds. The past light cone of any subsequent measurement point will be another such branching hypersurface for more and more new branches of the world. The complications that we encounter when there are intersecting past light cones of branchings are similar to the complications we encountered when there are intersecting past light cones of collapse (as in Fig. 5).

However, in the relativistic formulation of the Everett interpretation intersecting past light cones give rise to an extra complication in that the apexes of these light cones cannot be regarded as measurement points for one single observer. This is in sharp contrast to the nonrelativistic Everett interpretation, which is quintessentially solipsistic in that ever measurement can be viewed from the perspective of one single observer ("myself"). Relativistically a single-observer perspective is impossible, because the measurement points (that is, the apexes) of intersecting light cones have a spacelike separation, so one single observer cannot be present at two or more such apexes, and therefore the observers associated with different measurement points in Fig. 5 must be different. This modification of the radical solipsism of the Everett interpretation causes no difficulties of principle, because when one observer performs a measurement that triggers a branching of worlds along his past light cone, each of these branches will carry with it a copy of all the other observers, and we can presume that if any one of these other observers repeats the measurement her results will be consistent with what her colleague found on the first trial (this consistency of measurements by different observers is empirically true; but in an axiomatic formulation of a relativistic Everett scheme it may have to included as an explicit axiom? ).

In the Everett interpretation, as in the Copenhagen interpretation, the measurement begins with an interaction between the system and one or several detectors, and it ends with the observer's reception of data from the detectors and perception of the completed measurement. For a relativistic description of the measurement process, we have to assign to the observer's act of perception a specific point in space and time, where the world branches into many worlds (Everett) or where the wavefunction collapses, and from where this branching or collapse spreads progressively along the past light cone.

If we assume that the observer is a human observer who uses his eyes as detectors, we can gain clear intuitive understanding of why the past light cone plays a central role in the relativistic description of the measurement process. Suppose that the observer wishes to explore the probability distribution of a particle which is known to be in a finite region in his neighborhood. For the sake of simplicity, assume the particle is sufficiently large (or has a sufficiently large scattering cross section) so it can be seen with the available ambient light. The observer then merely scans the neighborhood by eye until he perceives the particle, and this completes the measuremen. Such a scanning measurement "by eye" is a natural instinctive behavior of humans—our evolutionary history has programmed us to always be on the lookout for desirable prey or for dangerous predators. A series of repetitions of this experiment, with identically prepared initial states, will give the observer the probability distribution for the particle.

However, the probability distribution measured in this way is not the probability distribution over a flat equal-time hypersurface. The light signal from the particle to the observer travels along the past light cone, so at the observer's eye a distant particle forms an image as it was a long time ago, and a nearby particle forms an image as it was a short time ago.[44] The complete picture that emerges from the combination of all experiments in the series shows the conditions along the past light cone, and the probability distribution deduced from the observer's visual observations is exactly the probability distribution corresponding to past-light-cone collapse (or past-light-cone branching of the world). It shows us our world as we see it whenever we scan our surroundings by eye.

Of course, this example does not provide a general proof of past-light-cone collapse. It merely deals with a special case—for a general proof we need the Schlieder argument. But, at least, this example illustrates that there is nothing bizarre at all about past-light-cone collapse. The only bizarre thing about past-light-cone collapse is that today, after more than a hundred years of relativity and discussions of quantum-measurements, hardly any physicists know about past-light-cone collapse. Ignorance is bliss?

A final important lesson to be learned from the relativistic analysis of collapse is that any proposal for new interpretation schemes must include an examination of the relativistic implications. By negligence (or by incompetence) proponents of new interpretation schemes blithely construct their machinery according to nonrelativistic quantum mechanics and hardly ever consider the relativistic aspects. The de Broglie-Bohm scheme of hidden variables and the Ghirardi-Rimini-Weber scheme with inclusion of random local collapses in the dynamical equations are glaring examples of this kind of negligence. In both of these schemes attempts at relativistic generalizations later showed that such generalizations are not viable, and it is disconcerting that neither the inventors of these schemes nor their credulous followers failed to recognize these defects immediately. Proponents of new interpretations need to keep in mind the obvious rule that every relativistically valid scheme has a Newtonian approximation, but not every conceivable Newtonian scheme can be generalized to a viable relativistic scheme.

## Acknowledgements

I thank Richard A. Mould for valuable comments on an early draft of this paper. And I thank Hagen Kleinert for bringing to my attention the deficiencies in the electron trajectories in Bohm's hidden-variables theory.[19]
My thinking about the problem of relativistic collapse was initially prompted by a puzzling example of collapse over astronomical distances posed by Roger Penrose in his provocative book *The Road to Reality*.[45]

## References and Endnotes

[1] A lively historical account of the development of the various interpretations of quantum mechanics from 1950 onward can be found in O. Freire Jr., *The Quantum Dissidents: Rebuilding the Foundations of Quantum Mechanics (1950-1990)* (Springer, Heidelberg, 2015). Although this books lacks technical details (there are hardly any equations in the entire book), its extensive lists of references are a good point of departure for readers who want to explore the quantum controversies.
[2] E. Wigner, "Interpretation of Quantum Mechanics," in J. A. Wheeler and W. H. Zurek, eds., *Quantum Theory of Measurement* (Princeton University Press, Princeton, 1983), p. 288.
[3] J. von Neumann, *Mathematical Foundations of Quantum Mechanics* (Princeton University Press, Princeton, 1955), p. 351, Eqs. (**1.**) and (**2.**).
[4] H. Everett III, Rev. Mod. Phys. **29**, 454 (1957). A good review of the Everett interpretation is given in the article "Everett's Relative-State Formulation of Quantum Mechanics" in the *Stanford Encyclopedia of Philosophy*; see https://plato.stanford.edu/entries/qm-everett/. Everett's interpretation was renamed "Many-Worlds" by DeWitt; see DeWitt and Graham, Ref. 32.
[5] Wigner speculated that "consciousness" of the observer plays a role in the completion of measurements. But I prefer Bohr's view, seconded by Wheeler, that measurements are completed by "an irreversible act of amplification" that renders the result "unambiguously communicable" (Wheeler and Zurek,, op. cit., p. 769).
[6] Wigner, op. cit., p. 287.
[7] E. Joos and H. D. Zeh, Zeitschrift Phys. B **59**, 223 (1985).
[8] W. H. Zurek, Phys.Today, October 1991, p. 36.
[9] R. Omnès, *The Interpretation of Quantum Mechanics* (Princeton University Press, Princeton, 1994); *Understanding Quantum Mechanics* (Princeton University Press, Princeton, 1999)
[10] M. Tegmark, Found. Phys. Lett. **6** (6), 571 (1993).
[11] A precursor of the decoherence interpretation was introduced by Bohm in 1951 in an example involving measurement of spin; see D. Bohm, *Quantum Theory* (Prentice-Hall, Englewood Cliffs, 1951), Sections 22.8-22.14. But this contribution of Bohm was forgotten when he quickly abandoned it in favor of a hidden-variable theory (see Ref. 17), where he used the Schrödinger wavefunction to construct a supplemental potential well that alters what is otherwise the motion of a classical fluid. However, this scheme was found to fail in relativity, when the Schrödinger equation is replaced by the Klein-Gordon equation (see Ref. 19). Despite this fatal defect, the scheme still has some stubborn adherents.
[12] S. Weinberg, *Lectures on Quantum Mechanics,* Cambridge University Press, Cambridge (2013), p. 87.
[13] Zurek, op. cit.
[14] S. L. Adler, Stud. Hist. Phil. Mod. Phys. **34**, 135 (2003).
[15] R. B. Griffiths, J. Stat. Phys. **36**, 219 (1984); see also *Consistent Quantum Theory* (Cambridge University Press, Cambridge, 2002).
[16] Wigner, op. cit., Eq. (42).
[17] D. Bohm, Phys. Rev. D **85**, 166, 180 (1965).
[18] G. C. Ghirardi, A. Rimini, and T. Weber, Phys. Rev. D **34**, 470 (1986).
[19] Bohm's hidden variable theory cannot be implemented for the relativistic Klein-Gordon equation, because the probability density $\propto i(\Phi^*\partial_0\Phi - \Phi^*\partial_0\Phi)$ is not positive definite, a fact that enthusiasts of de Broglie-Bohm mechanics obstinately shrug off ; see, e.g., P. R. Holland: *The Quantum Theory of Motion: An Account of the De*

*Broglie-Bohm Causal Interpretation of Quantum Mechanics* (Cambridge University Press, Cambridge, 1993). And even in the nonrelativistic regime there is clear empirical evidence against this theory in that a detailed calculation of the interference pattern produced by a beam diffracted by two slits leads to an intensity distribution noticeably different from that of the Fraunhofer pattern calculated from standard quantum mechanics and verified by experiments with electrons, neutrons, and atoms; see P. Chen and H. Kleinert, Electronic Journal of Theoretical Physics (EJTP) **13** (35), 1 (2016). As regards the Ghirardi-Rimini-Weber theory, an attempt at constructing a relativistic version of this theory showed that negative-energy wavefunctions—that is, antiparticles—would be generated whenever particles propagating through empty space collapse (or "flash") spontaneously; see R. Tumulka, arXiv:quant-ph/0406094 (2004).

[20] S. Weinberg, op. cit., p. 95.

[21] The most recent 2004 update of A. Cabello's compendium of 10,000+ references on the foundations of quantum mechanics and the interpretation of measurements (*Bibliographic guide to the foundations of quantum mechanics,* arXiv:quant-phys/0012089v12) lists only about twenty references in its specialized topical section on relativistic collapse.

[22] I. Bloch, Phys. Rev. **156**, 1377 (1967), Section III.

[23] S. Schlieder, Commun. Math. Phys. **7**, 305 (1968).

[24] K.-E. Hellwig and K. Kraus, Phys. Rev. D **1**, 566 (1970).

[25] Y. Aharonov and D. Z. Albert, Phys. Rev. D **21**, 3316 (1980); Phys. Rev. D **24**, 359 (1981); Phys. Rev. D **29**, 228 (1984). Several later papers by Aharonov and Albert on measurements of quantum fields are blighted by their rejection of past-light-cone collapse.

[26] To appreciate the daftness of this suggestion, the reader is invited to recall that different choices of Lorentz frames are merely different ways of choosing coordinates for spacetime. In a transformation of the vacuum state from an inertial to a *noninertial* reference frame, particles can be created (this is closely related to the Hawking radiation effect for black holes), but not in a transformation from one inertial reference frame to another.

[27] For the purposes of this analogy we assume that currents can flow even in regions where the (net) charge density is zero, as in the case of a current flowing on a neutral metallic wire. The "effective flow velocity" $\mathbf{v} \equiv \mathbf{j} / \rho$ can therefore be arbitrarily large and exceed the speed of light, which suggests there is no limit to the speed of propagation of probabilities.

[28] Note that, as an immediate consequence of our analogy, the probability density is not a scalar, but the 0 component of a four-vector current density. The nonrelativistic Schrödinger equation is misleading in that it suggests that the probability density is a scalar. Probability densities for the Klein-Gordon equation and the Dirac equation are 0 components of a conserved four-vector current density.

[29] Some definitions: Relative to the origin *O* of a Cartesian coordinate system in 4-D spacetime, the *nominal present* is the 3-D region $t = 0$, the *nominal future* is the region $t > 0$, and the *nominal past* is the region $t < 0$. The *future light cone* is the 3-D hypersurface $t-(x^2+y^2+z^2)^{1/2}=0$ and the *past light cone* is $t+(x^2+y^2+z^2)^{1/2}=0$. The *absolute future* is the region $t-(x^2+y^2+z^2)^{1/2}>0$, and the *absolute past* is the region $t+(x^2+y^2+z^2)^{1/2}<0$. There seems to be no generally accepted terminology for the region outside the light cones, that is, the region $x^2+y^2+z^2-t^2>0$. This is sometimes called the *elsewhere* region or the *spacelike* region, but Hellwig-Kraus call it the *outer cone*. The spacetime points in this region have an ambiguous time order relative to *O*, which can be changed to earlier or later by Lorentz transformations. Accordingly, I will call this the *ambiguous* region.

[30] For curved coordinates and/or curved spacetime, the flat spacetime volume element $d^3x$ must be replaced by the appropriate volume element $\sqrt{-g}d^3x$, which leads to corresponding replacements in the conservation law, which becomes $\partial_\mu(\sqrt{-g}\, j^\mu)=0$.

[31] This is analogous to what we do in taking a census of the total population of country, where we must include not only the population registered at fixed addresses, but also the migrant population of farmhands, tourists, circus performers, etc.

[32] B. DeWitt and N. Graham, *The Many Worlds Interpretation of Quantum Mechanics* (Princeton University Press, Princeton, 1973), p. 178.

[33] Hereafter, I prefer to speak of collapse of the wavefunction rather than the more abstract state vector, because the focus of this discussion is the propagation of the collapse in space and time, which requires a spacetime representation of the state vector. In the case of collapse of a state vector for spin, isospin, hypercharge, etc. for a particle-like system located at one unique spatial point, there is no spacetime distribution that raises questions about the when and where of the collapse. If the system has no spatial extent, the distinction between state vector and wavefunction becomes irrelevant.

[34] These three steps are laid out clearly by Bloch, op. cit. Note, however, that Bloch is obviously wrong in insisting that the apparatus states must be orthogonal—they must be *distinguishable* (that is, they must avoid excessive overlaps), but they need not be exactly orthogonal. Bloch himself comments that the macroscopic nature of the apparatus introduces random phases that eliminate interference terms and make the apparatus states behave as though they were orthogonal, even if they actually are not.

[35] R. A. Mould, "A defense of Hellwig-Kraus reductions," arXiv: quant-ph/9912044v1 (1999).

[36] Aharonov and Albert insist "So it seems the reduction process *must* be instantaneous," Phys. Rev. D **24**, 365 (1981).

[37] O. Cohen and B. J. Hiley, Foundations of Physics **25**, 1669 (1995).

[38] A. Aspect et al., Phys. Rev. Lett. **49**, 1804 (1982).

[39] W. Tittel et al., Phys. Rev. Lett. **81**, 3563 (1998).

[40] H. C. Ohanian, arXiv:1311.5840v2[quant-ph]. The 2017 version (v2) of this paper corrects a mistake in the procedure proposed for detection of coherent emission by two dipoles in the original 2013 version (v1).

[41] M. S. Chapman et al., Phys. Rev. Lett. **75**, 3783 (1995).

[42] Chapman et al. did not make any effort to detect the radiation pattern of the scattered laser light directly, but instead deduced this pattern from the recoils suffered by the atoms during the fluorescent re-emission of the photons. The recoil motion of the atoms generated observable shifts in the positions of the atom interference fringes.

[43] In arXiv:1311.5840v2[quant-ph] I allocated 5 ns for the laser excitation of the atoms and 5 ns for the subsequent decay interval. A crude estimate suggests that this is not the optimal choice for maximum intensity of fluorescent radiation.

[44] I have argued elsewhere that because our observations are restricted to the past light cone and its interior, these regions of spacetime must be regarded as the "real" world of physics; see H. C. Ohanian, "The Real World and Space-Time," in V. Petkov, ed., *Relativity and the Dimensionality of the World* (Springer, Dordrecht, 2007).

[45] R. Penrose, *The Road to Reality* (Knopf, New York, 2005), pp. 606-7.